\documentclass[aps,prc,showpacs,a4paper,nofootinbib,preprint]
{revtex4-1}
\usepackage{graphicx}
\usepackage{epsfig}
\usepackage{amssymb,latexsym,amsmath}
\usepackage{color}
\usepackage{hyperref}
\newcommand{\bs}{\mathbf}
\newcommand*{\dis}{\displaystyle}

\begin{document}

\title{Time dependence of
partition into spectators and participants
in
relativistic heavy-ion collisions}

\author{V. Vovchenko$^{1,2,3}$, D. Anchishkin$^{4,1}$, L.P. Csernai$^5$}

\affiliation{
$^1$Taras Shevchenko Kiev National University, Kiev 03022, Ukraine}
\affiliation{
$^2$Frankfurt Institute for Advanced Studies, Frankfurt 60438, Germany}
\affiliation{
$^3$Johann Wolfgang Goethe University, Frankfurt 60325, Germany}
\affiliation{
$^4$Bogolyubov Institute for Theoretical Physics, Kiev 03680, Ukraine}
\affiliation{
$^5$Institute for Physics and Technology, University of Bergen,
5007 Bergen, Norway}

\date{\today}


\begin{abstract}
The process of formation of the participant system in heavy-ion collisions
is investigated in the framework of a
simplified analytic Glauber-like model, which is based on the
relativistic Boltzmann
transport equation. The key point lies in the time-dependent partition
of the nucleon system into two groups: nucleons, which did not
take part in any interaction
before a given time and nucleons, which already have interacted.
In the framework of the proposed model we introduce a natural energy-dependent
temporal scale $t_c$, which allows us to remove all dependencies of the model
on the collision energy except for the energy dependence
of the nucleon-nucleon cross-section.
By investigating the time dependence of the total number of participants
we conclude
that the formation process of the participant system becomes complete
at $t\simeq1.5 t_c$.
Time dependencies of participant total angular momentum and vorticity
are also considered
and used to describe the emergence of rotation in the reaction plane.
\end{abstract}

\maketitle

\section{Introduction}
From the very beginning of the collision of two nuclei
some of the nucleons start to experience collisions and become
participants. The number of nucleons which have experienced collisions increases with time
and the number of the nucleons which did not take part in collisions decreases.
Finally, this results in the partition of
the total initial system of nucleons into two
subsystems: participants and spectators.
In the framework of the Glauber 
model~\cite{GlauberSitenko,BialasNWM,GlauberModel} (optical limit)
one can obtain average transverse distributions 
of the participants and spectators
at the end of this partition stage.  These
smooth distributions have been
used earlier as input to fluid dynamical models, 
see e.g., Refs.~\cite{Kolb2001,Kolb2001elliptic}.
The Monte Carlo Glauber (MC-Glauber) approach allows one to simulate
the initial partition stage on an event-by-event level
and can be used for determining fluctuating 
initial conditions in event-by-event
hydrodynamics~\cite{Holopainen2011,Schenke2011,Qiu2011}.
Fluctuations in the collective flow coefficients have been attributed to
initial spatial fluctuations~\cite{Broniowski2007,Alver2010}
and thus can be used to put constraints on the initial-state
geometry~\cite{Retinskaya2014,Renk2014}.
On the other hand, fluctuations can develop dynamically during
the fluid dynamical motion, especially if the matter undergoes 
a phase transition \cite{K1,K2,K3}.
While the transverse plane distribution (and its
fluctuations) of the formed participant system
has been investigated in literature in great detail by using the Glauber
approach, little attention was
paid to the temporal dynamics of the spectator-participant partition.
This
dynamics can be of special interest in peripheral
collisions where one can study, for instance, the process
of how participants gain a non-zero total angular momentum,
which in turn results in the emergence of
initial rotation in the reaction plane.
In the present work we develop an analytical Glauber-like model in the
framework of the relativistic Boltzmann equation (Sec. \ref{sec:model}) and use it
for the description of
the process of partition
into spectator and participant subsystems. Calculations done in the model
for various time-dependent quantities
are presented in Sec. \ref{sec:calculations} and 
conclusions are given in Sec. \ref{sec:conclusions}.

\section{The model}
\label{sec:model}

\subsection{Initial conditions and the ballistic mode}
In the simplest approximation
of our description within
the relativistic Boltzmann equation
we assume a ballistic mode, i.e., we neglect all the
reactions between hadrons and we separate the total system of net nucleons
into nucleons of the target (A) and projectile (B) nuclei.
The initial single-particle distribution functions 
$f^{(0)}_{A}(x,p)$ and $f^{(0)}_{B}(x,p)$ 
[hereinafter denoted $f^{(0)}_{A(B)}(x,p)$] of nucleons from corresponding nuclei
are described by the collisionless field-free relativistic Boltzmann equation
\begin{equation}
p^{\mu}\partial_{\mu}f^{(0)}_{A(B)}(x,p)\, =\, 0 \,.
\end{equation}
The solution to this equation is
\begin{equation}
f^{(0)}_{A(B)}(x,p)\ =\ \mathcal{F_{A(B)}}
\left[\bs r - \bs v (t-t_0),p\right]\,,
\label{eq:fABF}
\end{equation}
where $\mathcal{F_{A(B)}}(\bs r,p; t_0)$
is the distribution function of nucleons at the
initial time, $t_0$, $\bs v=\bs p/E_p$ is the velocity of particles and
$E_p=(m^2+\bs p^2)^{1/2}$.
We adopt the system of units $c=\hbar=1$.
The initial time, $t_0$, corresponds to the moment before any
interaction takes place.
I.e. no collision and no internal change within the two nuclei occurs
between $t=-\infty$ and $t_0$.

We assume that the initial distribution function of nucleons in
the  nucleus can be presented as
a product of a spatial and momentum distributions
\begin{equation}
\mathcal{F_{A(B)}}(\bs r,p; t_0)\, = \,
\rho_{A(B)}(\bs r; t_0)\ g_{A(B)}(\bs p)\ .
\label{eq:Frhog}
\end{equation}
Here $\rho_{A(B)}(\bs r; t_0)$ is the initial spatial distribution
of nucleons in the
target (projectile), and $g_{A(B)}(\bs p)$ is the initial
momentum distribution.
Since the collider center-of-mass (c.m.) frame and the Local Rest (LR) frame of a nucleus
are connected
via the Lorentz transformation in $(t,z)$ variables,
we can write the initial spatial density, $\rho_{A(B)}(\bs r; t_0)$,
(which is the 0th component of the nucleon 4-flow)
in the collider c.m. system (c.m.s.) in terms of corresponding 4-flow quantities
in terms of the Local Rest frame of the nucleus as
\begin{equation}
\rho_{A(B)}(\bs r; t_0) \, =\,
\gamma_0 \left\{ \rho_{A(B)}^{LR} [x, y, \gamma_0 (z - v_{A(B)} t_0)]
+
v_{A(B)} \, j_{z}^{A(B),LR} [x, y, \gamma_0 (z - v_{A(B)} t_0)] \right\} \,,
\label{eq:rhoinit}
\end{equation}
where $v_A = -v_B = v_0$ is the initial nucleus velocity in the c.m. frame,
$\gamma_0 = (1-v_0^2)^{-1/2}$,
$\rho_{A(B)}^{LR} (x,y,z)$ is the initial spatial distribution of nucleons
in the Local Rest Frame of the target (projectile) nucleus,
and $j_{z}^{A(B),LR}(x,y,z)$ is a $z$-coordinate of nucleon flow
in the same Local Rest Frame.

For the spatial distribution in the LR frame of the nucleus we
use the Woods-Saxon density profile
so that
\begin{equation}
\rho_{A(B)}^{LR} (x,y,z) \, = \, \rho_{_{WS}} (x \mp b/2, y, z) =
c_{\rho}\left\{1 + \exp\left[
\frac{\sqrt{(x \mp b/2)^2 + y^2 + z^2}-R_0}{a} \right] \right\}^{-1},
\end{equation}
where $a=0.545$~fm and $R_0$ is the nuclear radius.
The normalization constant $c_\rho$ is determined from the relation
$\int d \bs r \rho_{_{WS}}(\bs r) = A$, where $A$ is the mass number
of the nucleus.
In the above equation we have already taken into account a
shift in the $x$ coordinate due to the
non-zero impact parameter $b$.
It should be noted that our approach is not
restricted just to the standard Woods-Saxon profile,
other nuclear density profiles, i.e., three-parameter
Woods-Saxon, can also be used.
 Assuming that the momentum distribution of nucleons
in the LR frame of the nucleus is isotropic, we get that the particle flow
$j_{z}^{A(B),LR}$ vanishes, and the initial density,
 $\rho_{A(B)}(\bs r; t_0)$, in the collider c.m. frame
can be written as
\begin{equation}
\rho_{A(B)}(\bs r; t_0) \, =\, \gamma_0 \,
\rho_{_{WS}} [x \mp b/2, y, \gamma_0 (z - v_{A(B)} t_0)] \, .
\label{eq:rhoinit2}
\end{equation}

Expression~\eqref{eq:rhoinit2} corresponds to nuclear density in the moving
frame which has correct normalization, i.e., $\int d \bs r \, \rho_{A(B)}(\bs r; t_0) = A$.
To define the initial momentum distribution in the c.m.
frame we neglect the random Fermi motion in comparison to
the collective motion since we are dealing
with ultra-relativistic collision energies.
In this case the initial momentum distribution, $g_{A(B)}(\bs p)$, reads as
\begin{equation}
g_{A(B)}(\bs p)\ =\ \delta^2 (\bs p_{\bot})\,
\delta \left( p_z - p_{A(B)} \right),
\label{eq:gAB}
\end{equation}
where $p_A$ ($p_B$) is the initial momentum of nucleons
in the target (projectile).

Finally, we write the initial distribution function,
 $\mathcal{F_{A(B)}}(\bs r,p; t_0)$, as
\begin{equation}
\mathcal{F_{A(B)}}(\bs r,p; t_0)\, = \,
\gamma_0 \, \rho_{_{WS}} [x \mp b/2, y,
\gamma_0 (z - v_{A(B)} t_0)] \, \delta^2 (\bs p_{\bot})\,
\delta \left( p_z - p_{A(B)} \right)\ .
\label{eq:Frhog2}
\end{equation}
We can see
that the target and projectile initially move with opposite velocities
and they are completely separated spatially
at $t=t_0$, therefore indicating that the
presented initial conditions are consistent with the
condition that there are no reactions before the initial time $t_0$.

It can be seen that, in this particular case of
momentum distribution \eqref{eq:gAB}, the expression 
\eqref{eq:Frhog2} actually represents a
solution of the collision-less Boltzmann equation
if we treat $t_0$ as the time variable.
Indeed, using relation \eqref{eq:fABF} we can write the
time-dependent ballistic nucleon distribution functions in collider c.m. as
\begin{eqnarray}
f^{(0)}_{A(B)}(t,\bs r, \bs p) & = &
\gamma_0 \, \rho_{_{WS}} (x \mp b/2, y, \gamma_0
[z - \frac{p_z}{E_p} (t-t_0) - v_{A(B)} t_0]) \,
\delta^2 (\bs p_{\bot})\, \delta \left( p_z - p_{A(B)} \right)
\nonumber
\\
& = & \gamma_0 \, \rho_{_{WS}} (x \mp b/2, y, \gamma_0
[z - v_{A(B)} t]) \, \delta^2 (\bs p_{\bot})\,
\delta \left( p_z - p_{A(B)} \right)
\nonumber
\\
& = &\frac{\gamma_0 \, c_{\rho} \, \delta^2 (\bs p_{\bot}) \,
\delta \left( p_z - p_{A(B)} \right)} { 1 + \exp \left\{ \frac{\dis 1}{\dis a}
\left[ \sqrt{ (x \mp b/2)^2 + y^2 +
\gamma_0^2 \left( z - v_A t \right)^2 }
- R_0 \right] \right\} },
\label{eq:fAB}
\end{eqnarray}
where $E_p \equiv p^0$ is the energy of particle with four-momentum $p$ and
$p_z / E_p = v_z$.

\begin{figure}[h!]
    \begin{center}
     \includegraphics[width=.19\textwidth]{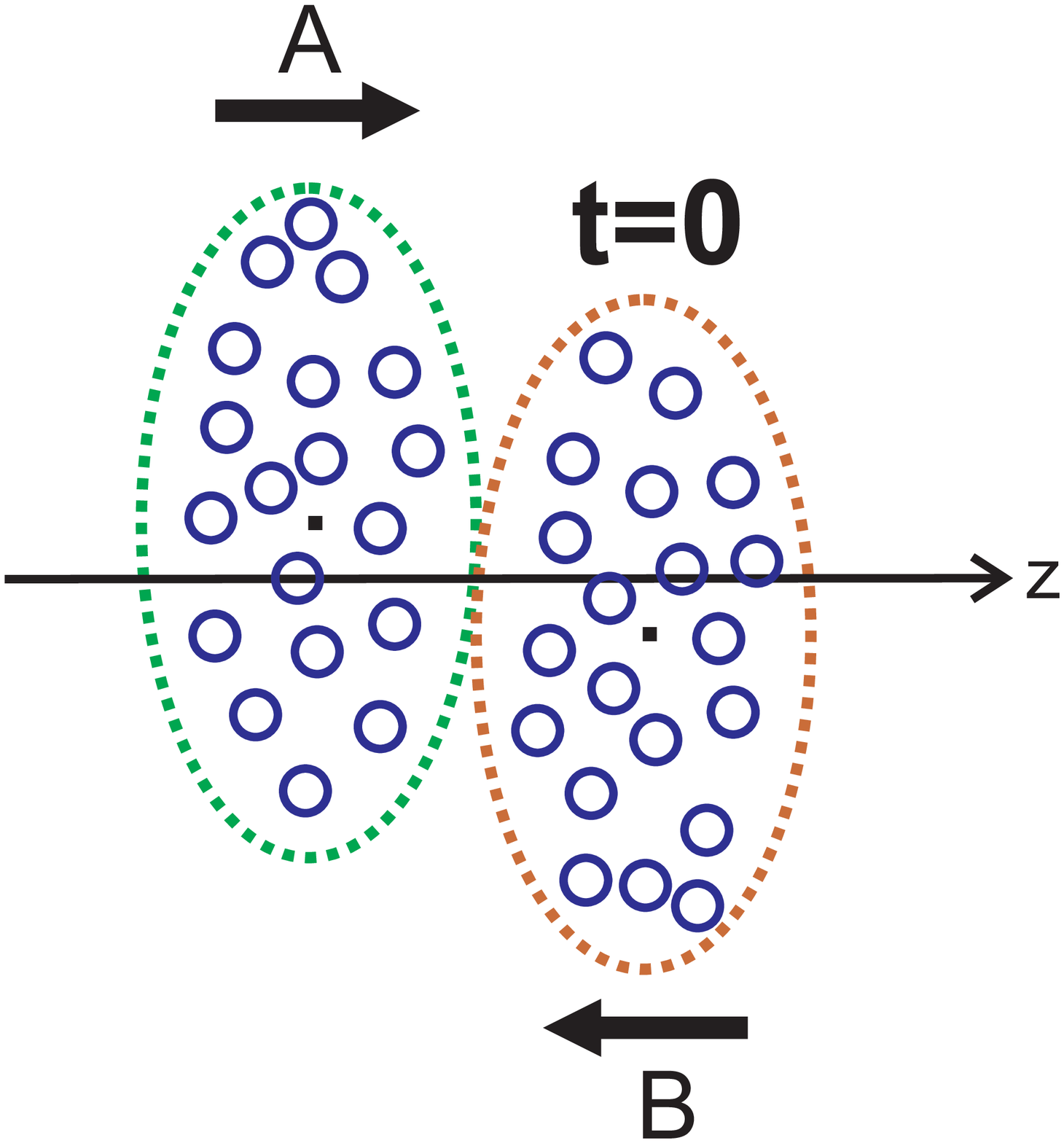}
     \includegraphics[width=.19\textwidth]{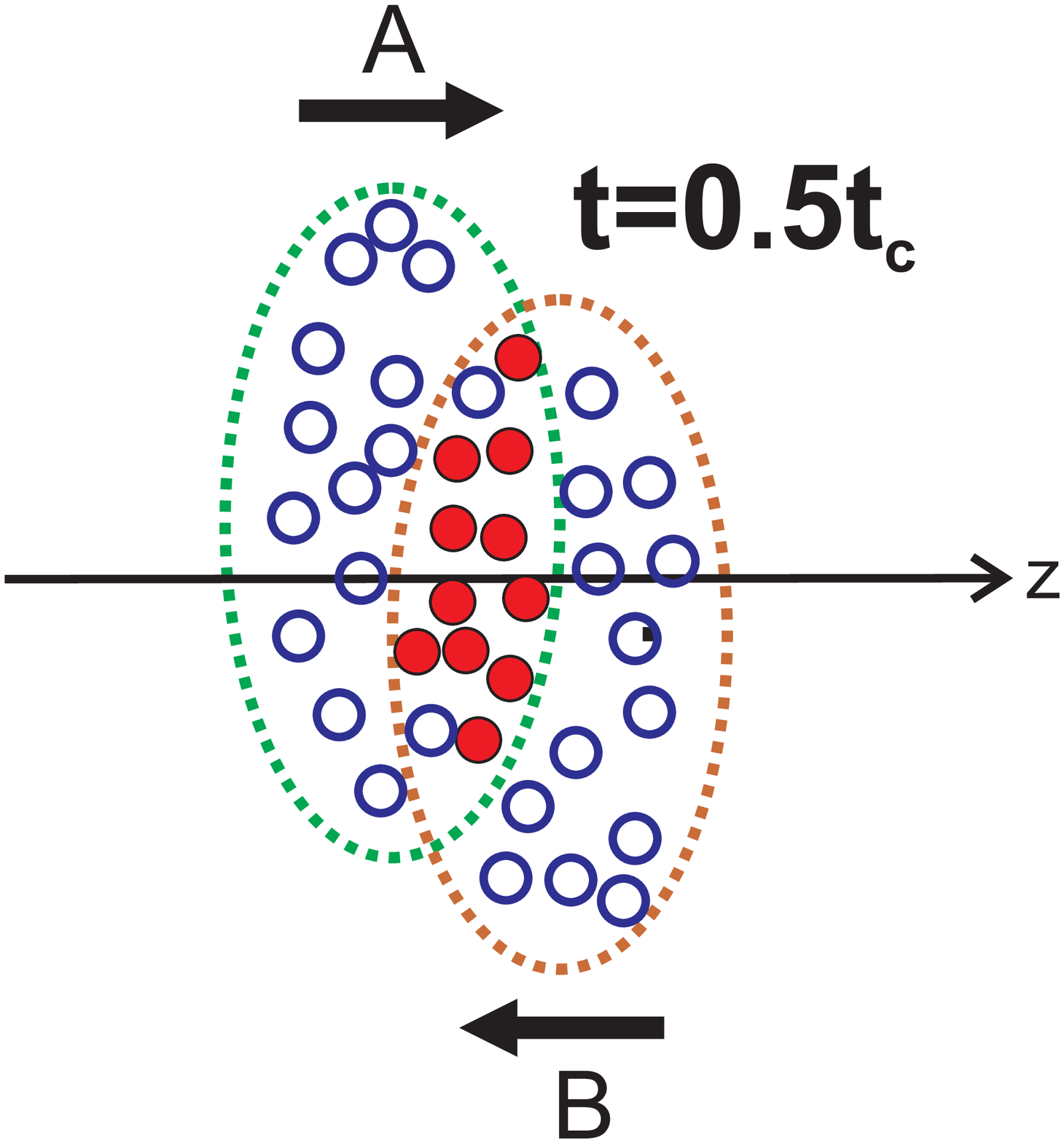}
     \includegraphics[width=.19\textwidth]{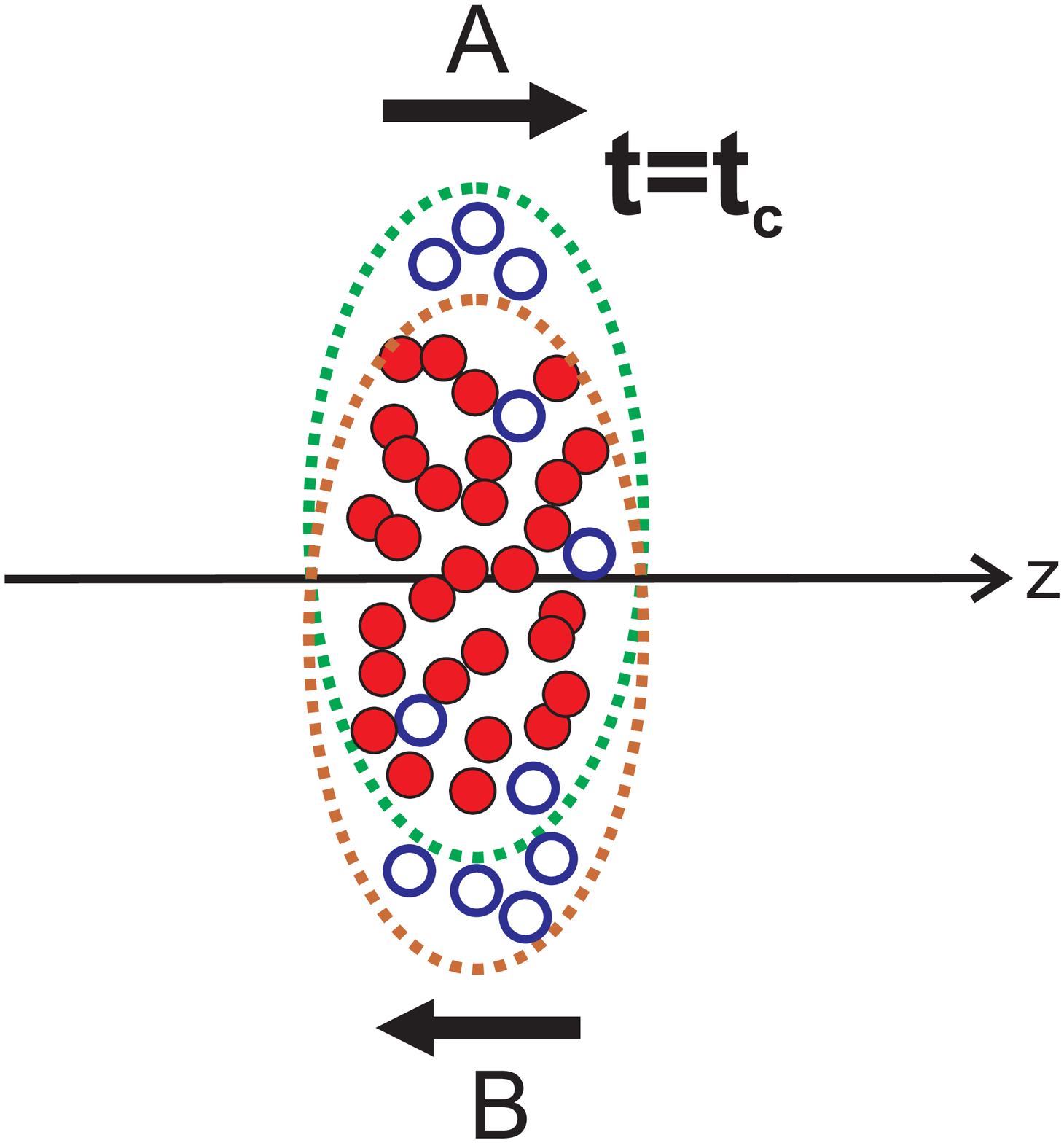}
     \includegraphics[width=.19\textwidth]{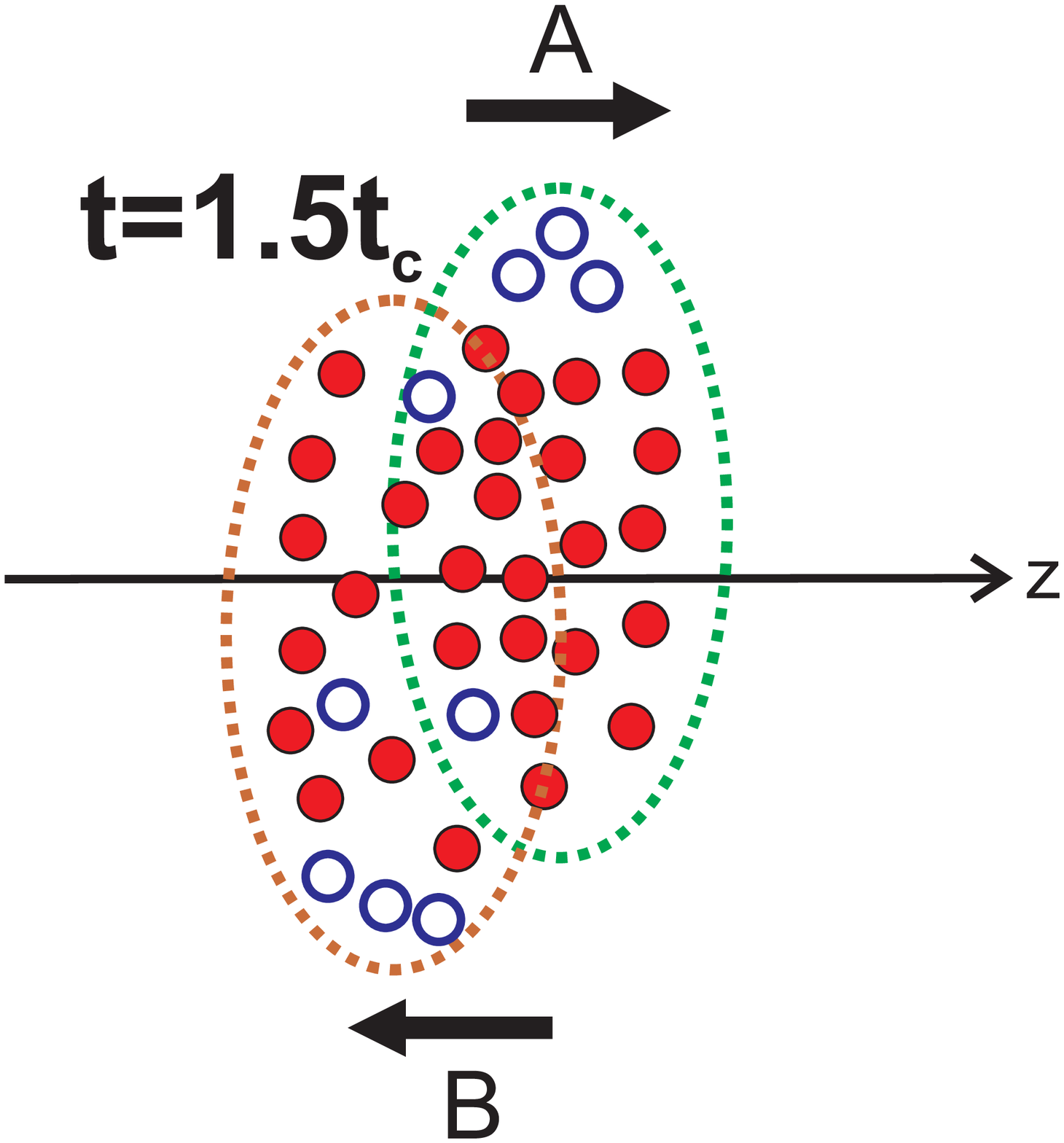}
     \includegraphics[width=.19\textwidth]{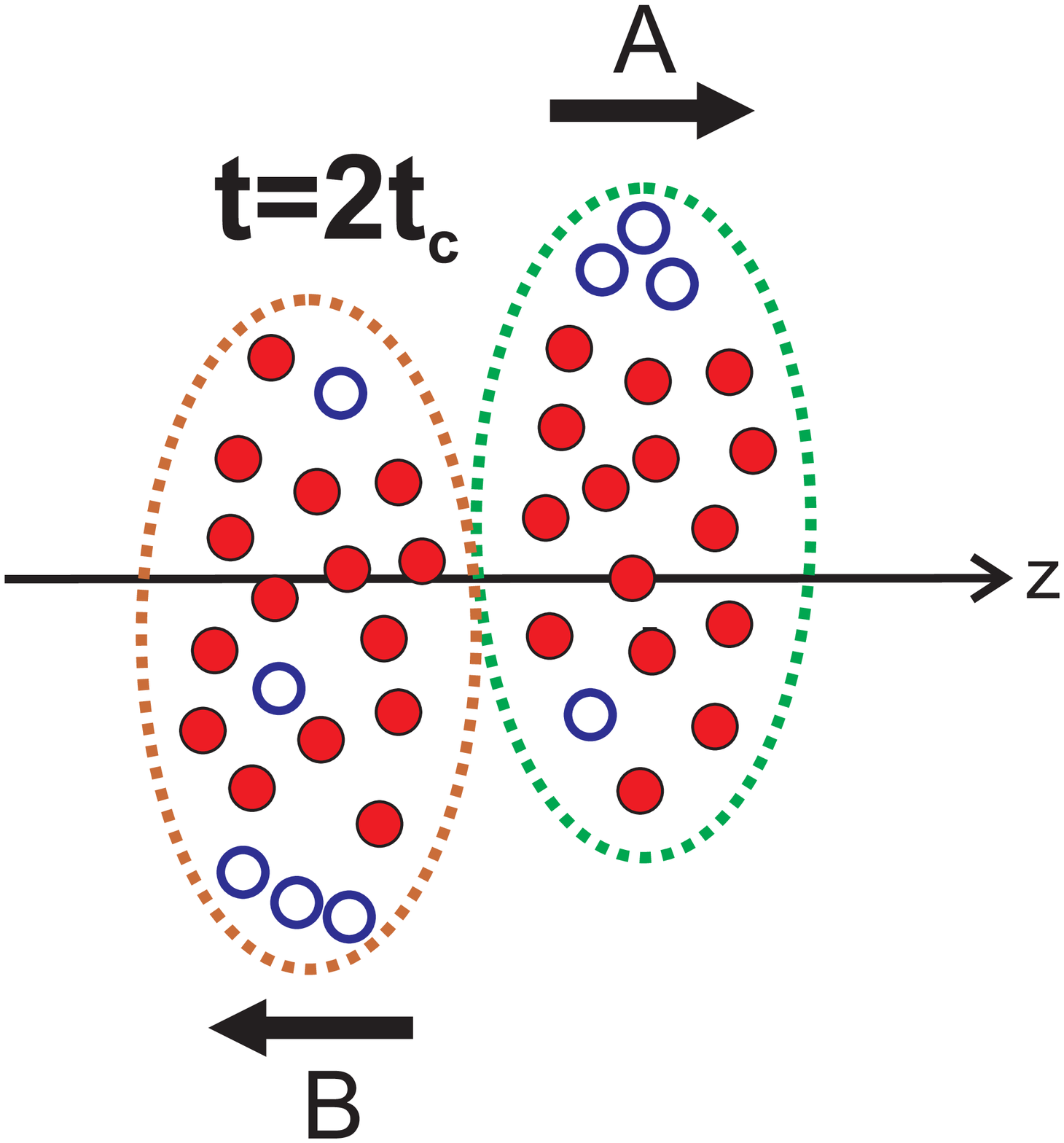}
     \caption{Schematic drawing of the system evolution in the proposed model.
     Blue points indicate nucleons which have not interacted before present time moment
     while red points indicate nucleons which already have interacted.}
     \label{pic:BallisticFireball}
    \end{center}
\end{figure}

It can be shown that the solution of the Boltzmann transport equation,
$\eqref{eq:fAB}$, has precisely the same structure
as the initial condition $\eqref{eq:Frhog2}$.
The presented ballistic distribution function corresponds to a
uniform motion of a nucleus with a Woods-Saxon nuclear density profile
which is
Lorentz-contracted in $z$-direction.
At the time moment $t=0$, the colliding nuclei experience
maximum density overlap and the $z$-coordinates of
their centers coincide, and are equal to zero.
For better correspondence to cascade models,
it makes sense to employ a time axis where at time
$t=0$, we have the $z$-coordinates of the centers of the
colliding nuclei separated by their Lorentz-contracted diameter,
 $2R_0/\gamma_0$ (see Fig.~\ref{pic:BallisticFireball}).
In such a way, the time $t=0$ approximately corresponds to the
time when the first reactions start to take place.
For instance, in case of central collisions it means
that at $t=0$ the colliding nuclei ``touch'' each other.
The timescale introduced above yields for the
time of the maximum overlap
$t_c = R_0/(\gamma_0 \, v_0)$.
Consequently, we obtain the time-dependent ballistic nucleon
distribution functions
in their final form
\begin{eqnarray}
f^{(0)}_{A(B)}(t,\bs r, \bs p) & = &
\rho_{A(B)}^{(0)}(t, \bs r) \, \delta^2 (\bs p_{\bot}) \,
\delta \left( p_z - p_{A(B)} \right)
\nonumber
\\
& = &\frac{\gamma_0 \, c_{\rho} \, \delta^2 (\bs p_{\bot}) \,
\delta \left( p_z - p_{A(B)} \right)} { 1 + \exp \left\{ \frac{\dis 1}{\dis a}
\left[ \sqrt{ (x \mp b/2)^2 + y^2 +
\gamma_0^2 \left( z \pm R_0/\gamma_0 \mp v_0 t \right)^2 }
- R_0 \right] \right\} }\,,
\label{eq:fAB_final}
\end{eqnarray}
where $\rho_{A(B)}^{(0)}(t, \bs r) =
\gamma_0 \, \rho_{_{WS}} \left(x \mp b/2, y,
\gamma_0 [z \mp v_0 (t-t_c)]\right)$.

\bigskip

\subsection{Partition into spectators and participants}
In this section we describe the process of partition 
of nucleons into
spectators and participants.
We assume that nucleons
coming from
the target (projectile) become participants
in collisions with nucleons from projectile (target).
We define $f^{S}_{A(B)}(t,\bs r, \bs p)$ as the distribution function
of nucleons from the target (projectile), which had not
taken part in any reactions before time $t$ in the collider
c.m. frame. It is seen from the definition that, at $t \to \infty$, this
distribution function describes all spectators
in the collision.
Following this definition and also the above-mentioned assumption
about collisions where nucleons become participants, we can
describe the functions $f^{S}_{A(B)}(t,\bs r, \bs p)$
by the Boltzmann transport equation by assuming
binary collisions, local molecular chaos, and collision integrals
containing only ``loss'' terms.
For instance, for nucleons from the target we have
\begin{equation}
p^{\mu} \partial_\mu f_A^{S} (t,\bs r, \bs p)
= -\frac{1}{2}\int \frac{d^3 p_1}{E_{p_1}} \,
\frac{d^3 p'}{E_{p'}} \frac{d^3 p'_1}{E_{p'_1}}
f_{A}^{S}(t, \bs r, \bs p) \, f_{B}^{(0)}(t, \bs r, \bs p_1)
W(p,p_1 | p', p_1'), \label{eq:fA1}
\end{equation}
where $W(p,p_1 | p', p_1')$ is the transition rate.

In order to perform integrations in Eq. \eqref{eq:fA1} we will use the
transition rate
$W(p,p_1 | p', p_1') = s \, \sigma(s, \theta) \,
\delta^4(p+p_1-p'-p_1')$
for elastic binary collisions, where $s \equiv (p+p_1)^2$
and $\sigma(s, \theta)$ is the differential cross section of nucleon-nucleon collision.

Since we are only considering ``loss'' terms, only
the total nucleon-nucleon cross section
will be relevant for the final result.
After integrating \eqref{eq:fA1}
over outgoing particle momenta $p'$ and $p'_1$ we get
\begin{equation}
p^{\mu} \partial_\mu f_A^{S} (t,\bs r, \bs p) =
 -\frac{1}{2}\int \frac{d^3 p_1}{E_{p_1}} d \Omega \sigma(s,\theta)
\frac{1}{2} \sqrt{s (s-4m^2)} f_{A}^{S}(t, \bs r,
\bs p) \, f_{B}^{(0)}(t, \bs r, \bs p_1).
\label{eq:fA2}
\end{equation}

Taking into account that
$\displaystyle \frac{1}{2}
\int d \Omega \sigma(s, \theta) = \sigma_{_{NN}}(s)$
and using explicit expression for $f_{A}^{(0)}$ \eqref{eq:fAB_final}
we perform the integration over $p_1$
\begin{equation}
p^{\mu} \partial_\mu f_A^{S} (t,\bs r, \bs p)
= - \frac{\sigma_{_{NN}}(s)}{E_{p_0}}
\frac{1}{2} \sqrt{s (s-4m^2)} f_{A}^{S}(t, \bs r, \bs p) \,
\rho_{B}^{(0)}(t, \bs r).
\end{equation}
Since $f_A^{S} (t,\bs r, \bs p)$
describes nucleons, which did not take part in
any reactions, it can be expressed as
\begin{equation}
f_{A}^{S}(t, \bs r, \bs p)
= \rho_A^{S} ( t, \bs r) \, \delta^2(\bs p_{\perp}) \, \delta(p_z-p_A),
\label{eq:fAdecom}
\end{equation}
where $p_A = -p_B = p_0$ and $\rho_A^{S} ( t, \bs r)$
is the time-dependent spatial density of the spectator nucleons.
Then, taking into account that
$E_{p_0} = \displaystyle \frac{\sqrt{s}}{2}$
and $p_0 = \displaystyle \frac{1}{2}(s-4m^2)^{1/2}$,
we get the equation for $\rho_A^{S} (t, \bs r)$
\begin{eqnarray}
p_0^{\mu} \partial_{\mu} \rho_A^{S} ( t, \bs r) & = &
-2 \sigma_{_{NN}} p_0 \rho_A^{S}(t, \bs r) \,
\rho_{B}^{(0)}(t, \bs r), \label{eq:rhoA1} \\
\rho_A^{S}(t_0, \bs r) & = & \rho_A^{(0)}(t_0, \bs r).
\label{eq:rhoA1init}
\end{eqnarray}
Here the expression on the right-hand side of
Eq. \eqref{eq:rhoA1} is proportional
to the number of binary collisions in the
four-volume element at $(t, \bs r)$, between any nucleons from
target (B) and those nucleons from projectile (A), which had not yet interacted
at time $t$.
It is seen that this expression depends only on
spatial densities, relative velocity and the
nucleon-nucleon cross section.
Thus, if we regard $\sigma_{_{NN}}$ as the total
nucleon-nucleon cross section then
Eq.~\eqref{eq:rhoA1init} also describes
the loss of the non-interacting nucleons due to
any binary reactions of nucleons and not just
due to elastic collisions.
The solution of Eq. \eqref{eq:rhoA1}
with initial condition \eqref{eq:rhoA1init}
can be written as
\begin{equation}
\rho_A^{S} ( t, \bs r) = \rho_A^{(0)}(t, \bs r) \,
\exp \left\{ -2 \sigma_{_{NN}} v_0  \int_{t_0}^{t}
d t'\rho_{B}^{(0)}[t', \bs r - \bs v_A (t-t')]\right\},
\end{equation}
where $v_0 = p_0 / E_{p_0}$ and $\bs v_A = (0, 0, v_0)$.
Similarly, for nucleons from the projectile we have
\begin{equation}
\rho_B^{S} ( t, \bs r) = \rho_B^{(0)}(t, \bs r) \,
\exp \left\{ -2 \sigma_{_{NN}} v_0  \int_{t_0}^{t}
d t'\rho_{A}^{(0)}[t', \bs r - \bs v_B (t-t')]\right\},
\end{equation}
where $\bs v_B = (0, 0, -v_0)$.

\subsection{Transverse distribution of spectators}
It is easy to see similarities between our model and the optical limit
of the Glauber-Sitenko approach \cite{GlauberSitenko}
applied for the description of relativistic heavy-ion collisions.
Indeed, in our simplified kinetic approach we consider only binary
collisions between nucleons which
always move in the forward-backward direction, and the probability of
binary interaction
is determined by the total nucleon-nucleon cross section.
One of the quantities which can be evaluated in that approach is
the transverse distribution $T^{\rm part} (x,y)$ of the
wounded nucleons (participants) \cite{BialasNWM,GlauberModel}, which
is often used to define
initial conditions in fluid dynamical models assuming that
the transverse expansion of
the interacting system is small during the initial pre-equilibrium phase.
This distribution reads as
\begin{eqnarray}
& & T^{\rm part} (x,y)  = T_A^{\rm part} (x,y) + T_B^{\rm part} (x,y)
\nonumber \\
& & = T_A (x-b/2,y) \left[1 -
\left(1 - \frac{\sigma_{_{NN}} T_B (x+b/2,y)}{A}\right)^A\right]
\nonumber \\
& & \quad + T_B (x+b/2,y)
\left[1 - \left(1 - \frac{\sigma_{_{NN}} T_A (x-b/2,y)}{A}\right)^A \right]
\nonumber \\
& & \approx T_A (x-b/2,y) \,
\left[1 - \exp\left\{-\sigma_{_{NN}} T_B (x+b/2,y)\right\}\right] +
\nonumber \\
& & \quad T_B (x+b/2,y) \, \left[1 -
\exp\left\{-\sigma_{_{NN}} T_A (x-b/2,y)\right\} \right],
\label{eq:trpartGS}
\end{eqnarray}
where $T_{A(B)} (x,y) = \int dz \, \rho_{_{WS}} (x,y,z)$
is the nuclear thickness function (normalized to $A$).
Consequently, the transverse distribution of spectators
can be written as
\begin{eqnarray}
& & T^{\rm spec} (x,y)
= T^{\rm tot} (x,y) - T^{\rm part} (x,y) \nonumber \\
& & = T_A (x-b/2,y)
\left(1 - \frac{\sigma_{_{NN}} T_B (x+b/2,y)}{A}\right)^A
+ T_B (x+b/2,y) \left(1 - \frac{\sigma_{_{NN}} T_A (x-b/2,y)}{A}\right)^A
\nonumber \\
& & \approx T_A (x{-}b/2,y) \,
\exp\left\{-\sigma_{_{NN}} T_B (x{+}b/2,y)\right\}
+ T_B (x{+}b/2,y) \, \exp\left\{-\sigma_{_{NN}} T_A (x{-}b/2,y)\right\}.
\label{eq:trspecGS}
\end{eqnarray}

To make a quantitative comparison of our model with
the above-mentioned approach
we calculate the transverse distribution of spectators within our model.
To account for all possible nucleon interactions we let the initial time
moment $t_0 \to -\infty$.
Then the transverse distribution of spectators from projectile
$T_A^{\rm spec} (x,y)$
can be calculated as
\begin{eqnarray}
T_A^{\rm spec} (x,y) & = & \lim_{t \to \infty} \int d \bs p
\int dz  \, f_{A}^{S}(t, \bs r, \bs p) \nonumber \\
& = & \lim_{t \to \infty} \int dz \, \rho_A^{(0)}(t, \bs r) \,
\exp\left\{-2 \sigma_{_{NN}} v_0  \int_{-\infty}^{t}
d t' \rho_{B}^{(0)}[t', \bs r - \bs v_A (t-t')]\right\}.
\end{eqnarray}
To perform the integration in the exponent we use
$$
\rho_{B}^{(0)}[t', \bs r - \bs v_A \left(t-t'\right)]
= \gamma_0 \, \rho_{_{WS}} [x + b/2, y,
\gamma_0 (z - v_0 t + 2 v_0 t' - v_0 t_c) ]\ ,
$$
and make the transformation of the integration variable:
$\displaystyle t'
= \frac{1}{2 v_0 \gamma_0} \left[ z' - \gamma_0 z
+ v_0 \gamma_0 (t+t_c)\right]$.
By using that $t \to \infty$ and also the definition of the
nuclear thickness function we can
perform the integration over the new variable
$z'$ under the exponent and get
\begin{equation}
T_A^{\rm spec} (x,y) = \lim_{t \to \infty}
\int dz \, \rho_A^{(0)}(t, \bs r) \,
\exp\left\{-\sigma_0 T_B(x+b/2,y) \right\}.
\end{equation}
By using that $\int dz \, \rho_A^{(0)}(t, \bs r) = T_A (x-b/2,y)$
we finally get
\begin{equation}
T_A^{\rm spec} (x,y)
= T_A (x-b/2,y) \, \exp\left\{-\sigma_{_{NN}} T_B(x+b/2,y) \right\}.
\label{eq:TA}
\end{equation}
Similarly, the transverse distribution of spectators from projectile reads as
\begin{equation}
T_B^{\rm spec} (x,y) = T_B (x+b/2,y) \,
\exp\left\{-\sigma_{_{NN}} T_A(x-b/2,y) \right\}.
\label{eq:TB}
\end{equation}

Comparing Eqs. \eqref{eq:TA}-\eqref{eq:TB} with \eqref{eq:trspecGS}
we can conclude that our model is consistent with the Glauber-based approach
for describing heavy-ion collisions. Furthermore, it provides the
possibility of studying the time-dependent features of the
spectator-participant
partition process in the early stage of the nucleus-nucleus collision.
Comparison
of our model with MC-Glauber is presented in Appendix A.

\section{Calculation results}
\label{sec:calculations}
To study the temporal structure of the partition of spectators and
participants
we consider the time-dependent transverse distribution $T^{\rm s} (t;x,y)$
of the nucleons, which did not interact before time $t$.
This distribution reads
\begin{eqnarray}
T^{\rm s} (t;x,y) & = & T_A^{\rm s} (t;x,y) + T_B^{\rm s} (t;x,y), \\
T_{A(B)}^{\rm s} (t;x,y) &
= & \int d \bs p \int dz  \, f_{A(B)}^{S}(t, \bs r, \bs p) \nonumber \\
& = & \int dz \, \rho_{A(B)}^{(0)}(t, \bs r) \,
\exp\left\{-2 \sigma_{_{NN}} v_0  \int_{-\infty}^{t} d t'
\rho_{B(A)}^{(0)}[t', \bs r - \bs v_{A(B)} (t-t')]\right\}.
\end{eqnarray}
We can rewrite this expression in terms of the initial Woods-Saxon distribution:
\begin{eqnarray}
T_{A(B)}^{\rm s} (t;x,y) & = & \int dz \,
\gamma_0 \rho_{_{WS}}(x \mp b/2,y,\gamma_0[z \mp v_0 (t-t_c)]) \times
\nonumber \\
& & \quad \exp\left\{-2 \sigma_{_{NN}} v_0  \int_{t_0}^{t} d t'
\gamma_0 \rho_{_{WS}}(x \pm b/2,y,
\gamma_0 [z \mp v_0 (t{+}t_c) \pm 2 v_0 t'])\right\}\ .
\label{eq:TABt1}
\end{eqnarray}
It is useful to introduce the variables
$\tilde{z} = \gamma_0 z$ and
$\tilde{t} = t/t_c$,
where, as previously defined,
$t_c = R_0 / (\gamma_0 v_0)$,
is the
time of the maximum overlap of the colliding nuclei
(see Fig. \ref{pic:BallisticFireball}).
Studies within Monte Carlo cascade models
have shown that this time moment corresponds
to the maximum of the nucleon-nucleon collision
frequency \cite{Anchishkin2010,Anchishkin2013PRC,Anchishkin2013}
and it appears to be a natural energy-dependent
temporal scale for the initial stage of the collision.
This time, $t_c$, decreases with increasing
collision energy and lies in the range:
$t_c \simeq 1$ to $2$~fm/$c$ at energies of the CERN Super Proton Synchrotron (SPS),
$t_c \simeq 0.1$ to $0.8$~fm/$c$ at energies of the BNL Relativistic Heavy Ion Collider (RHIC) and
$t_c \sim 10^{-2}$ to $10^{-3}$~fm/$c$ at energies of the Large Hadron Collider (LHC).
Equation \eqref{eq:TABt1} is then rewritten as
\begin{eqnarray}
T_{A(B)}^{\rm s} (\tilde{t};x,y) & = & \int d \tilde{z}
\rho_{_{WS}}[x \mp b/2,y,\tilde{z} \mp R_0 (\tilde{t}-1)] \times
\nonumber \\
& & \quad \exp\left\{-2 \sigma_{_{NN}} R_0
\int_{-\infty}^{\tilde{t}} d \tilde{t}' \rho_{_{WS}}
[x \pm b/2,y,\tilde{z} \mp R_0 (\tilde{t}+1) \pm 2 R_0 \tilde{t}']
\right\}.
\label{eq:TABt2}
\end{eqnarray}

\subsection{Number of participants}

The total number of participants
(net-baryon participant number)
at time $t$ can be obtained as

\begin{equation}
N_{\rm part} (t)
= 2A - \int dx dy \left[ T_A^{\rm s}(t;x,y) + T_B^{\rm s}(t;x,y) \right].
\end{equation}

\begin{figure}
\begin{minipage}{.48\textwidth}
\centering
\includegraphics[width=\textwidth]{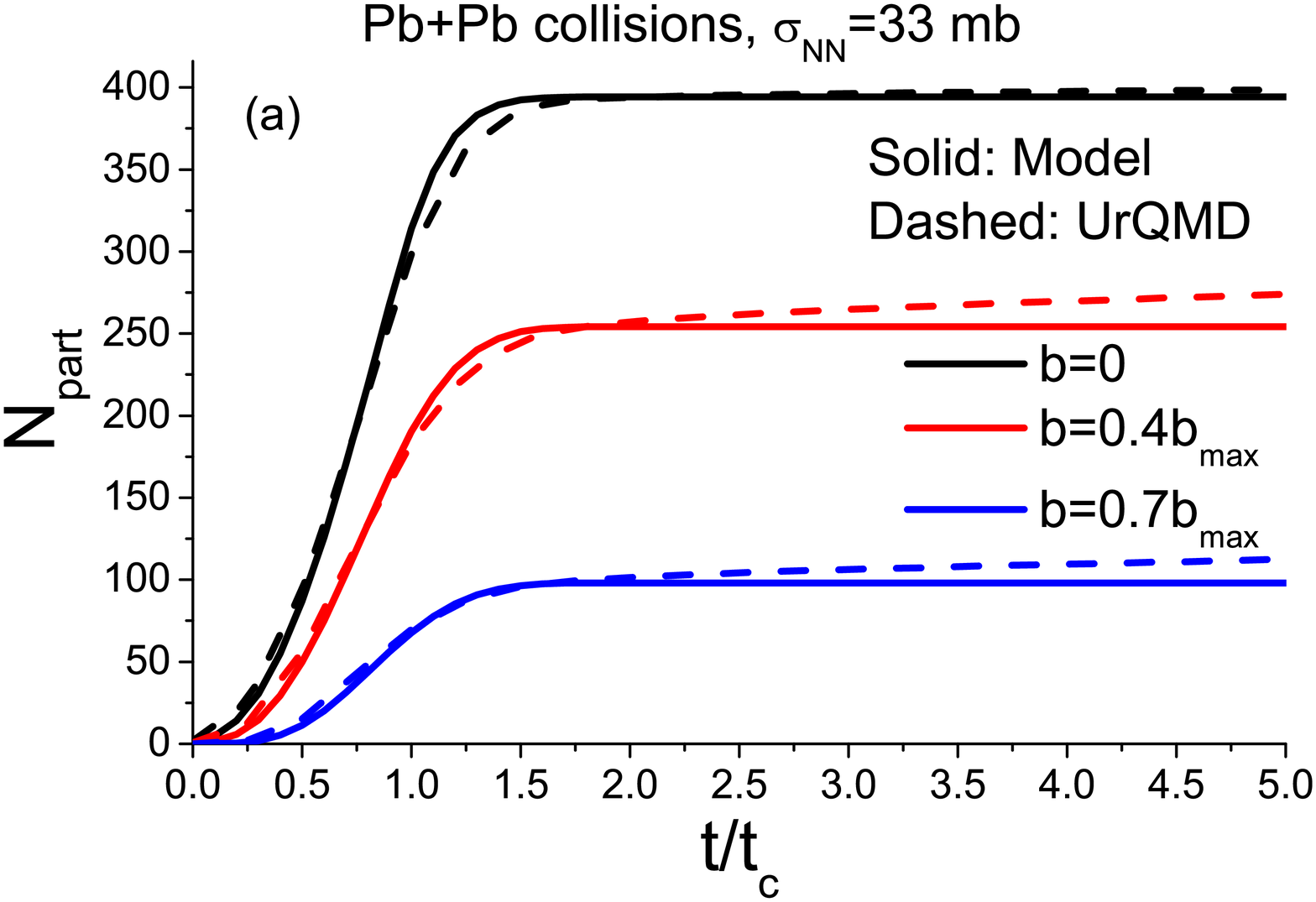}
\end{minipage}
\begin{minipage}{.48\textwidth}
\includegraphics[width=\textwidth]{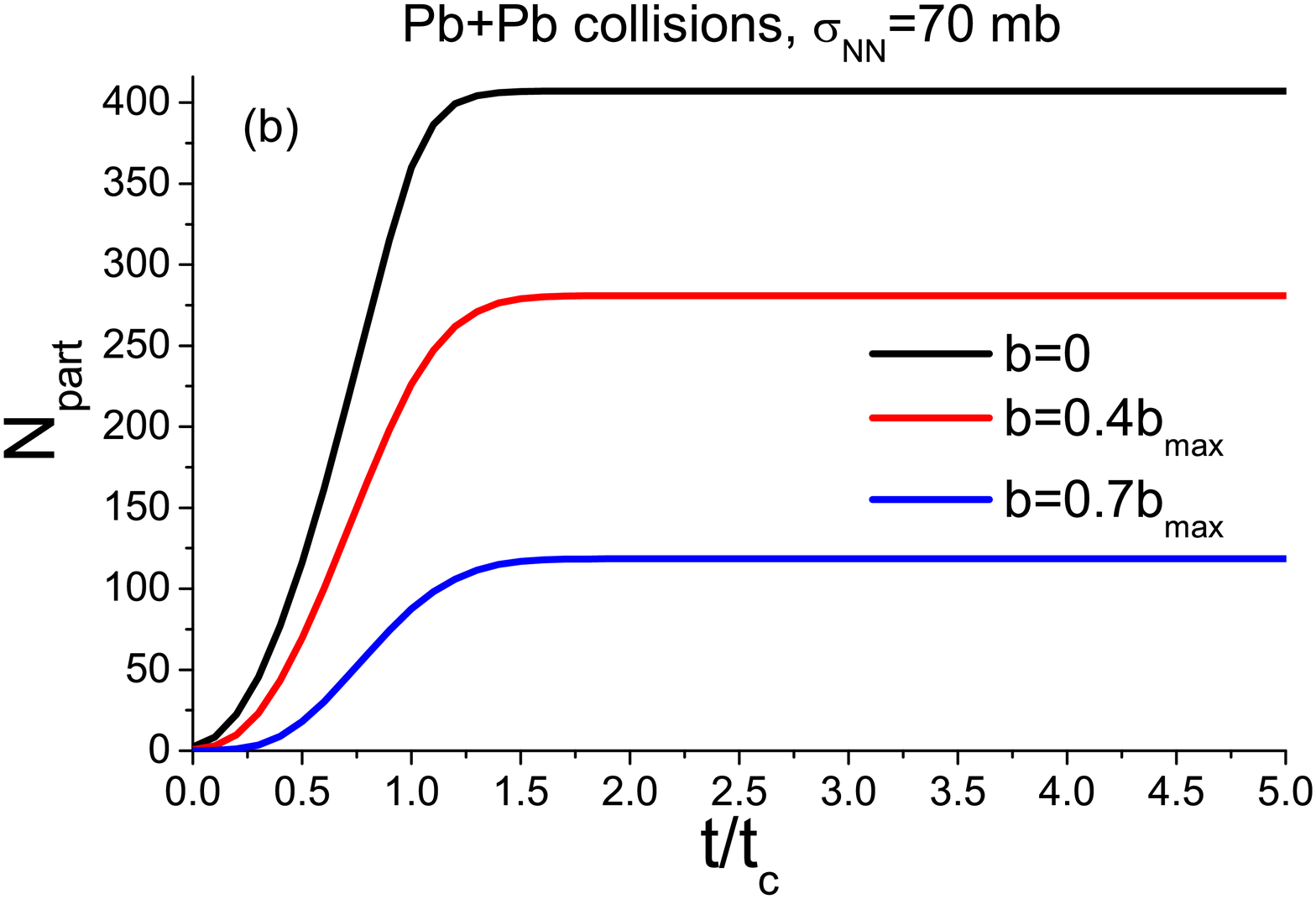}
\end{minipage}
\caption{The time dependence of the total number of participant nucleons
in Pb+Pb collisions at
(a) SPS and RHIC energies ($\sigma_{_{\rm NN}} = 33$~mb) and
(b) LHC energy ($\sigma_{_{\rm NN}} = 70$~mb)
for different values of impact parameter.
Solid lines depict calculations in the proposed model
while dashed lines in panel (a) correspond to calculations
from the UrQMD model at $\sqrt{s_{_{\rm NN}}}=17.3$~GeV.
}
\label{fig:Npart}
\end{figure}

The time dependence of the total number of participant nucleons
in Pb-Pb collisions is depicted in Fig.~\ref{fig:Npart} for
(a) SPS and RHIC energies ($\sigma_{_{\rm NN}} = 33$~mb) and
(b) LHC energy ($\sigma_{_{\rm NN}} = 70$~mb) at
three different centralities:
$b=0,0.4b_{\rm max},0.7b_{\rm max}$, where $b_{\rm max} = 2R_0$
and $R_0 = 6.53$~fm.
We can see that a change in the nucleon-nucleon cross section,
which roughly corresponds
to the increase of the collision energy from RHIC to LHC,
has little influence on the
time dependence of $N_{\rm part} (t)$ and only slightly increases
the total number of participant
nucleon charge at the given impact parameter.
It is seen from Fig.~\ref{fig:Npart} that the
formation of the participant
system is the most intense in the time range $t \simeq 0.5t_c$ to $1 t_c$ and
becomes complete at about $t=1.5 t_c$.

It makes sense to make a comparison of predictions
regarding time dependence of our simplified analytic model
with a more complicated cascade model
such as
the ultrarelativistic quantum molecular dynamics
(UrQMD) transport approach~\cite{UrQMD1998,UrQMD1999}.
The time dependence of the average total number of participant net nucleons (baryons)
can be calculated in UrQMD as event-by-event average of
$N_{\rm part} (t) = 2A - N_{\rm spec} (t)$,
where $N_{\rm spec} (t)$ is determined in each event
by analyzing the collision history.
UrQMD results for $N_{\rm part} (t)$ in Pb+Pb collisions
at top SPS energy of $\sqrt{s_{_{\rm NN}}}=17.3$~GeV
are depicted by dashed lines in Fig.~\ref{fig:Npart}a.
We note that 
the 
temporal axis in UrQMD is specially aligned
in Fig.~\ref{fig:Npart}a
with the one used in our model so that the
time moment $t=0$ correspond to two colliding nuclei
``touching'' each other.
The comparison of UrQMD with calculations of our model (solid lines in Fig.~\ref{fig:Npart}) shows
generally good agreement between our model and UrQMD.
One can see, however, that 
the 
number of participants in UrQMD keeps increasing, albeit insignificantly,
also at times $t>1.5t_c$, which can be attributed to the more complex collision dynamics
of UrQMD compared to our analytic model.

\subsection{Angular momentum}
Another important quantity, of which the time dependence
can be studied within the proposed model,
is the total angular momentum of the participant system.
The total angular momentum of the formed participant system
is non-zero in non-central collisions \cite{Becattini2008,Gao2008} and
can attain a significantly large value
($L \approx 10^6 \hbar$ for LHC energies \cite{Vovchenko2013}).
The angular momentum illustrates the initial
rotation of the system of participants, and
it was shown that it depends strongly on
the initial nuclear density profile and leaves some freedom
for the assumed initial state of the participant system in
fluid dynamical and
in molecular dynamics models.
The time-dependent total angular momentum, $L_{\rm tot}^P (t)$,
of the participant system
can be calculated in our model as the difference of total
angular momentum, $L_{\rm tot}$, and
the time-dependent angular momentum, $L_{\rm tot}^S (t)$,
of nucleons, which did not interact before time $t$.
These quantities can be written as
\begin{eqnarray}
L_{\rm tot} & = & p^z_{\rm in} \int dx dy \, x
\left[ T_A(x-b/2,y) - T_B(x+b/2,y) \right], \\
L_{\rm tot}^S (t) & = & p^z_{\rm in} \int dx dy \, x
\left[ T_A^S(t;x,y) - T_B^S(t;x,y) \right], \\
L_{\rm tot}^P (t) & = & L_{\rm tot} - L_{\rm tot}^S (t),
\end{eqnarray}
where $p^z_{\rm in} = (s/4 - m_{N}^2)^{1/2}$
is the initial momentum of a nucleon.

\begin{figure}
\begin{center}
\includegraphics[width=0.65\textwidth]{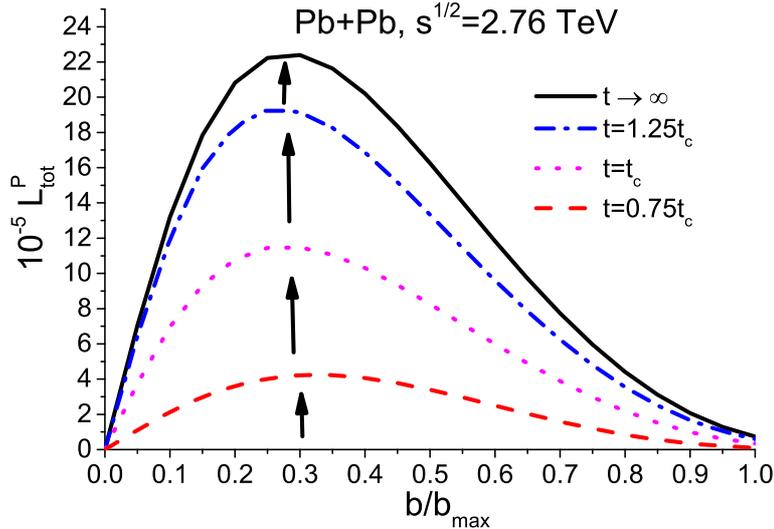}
\caption{The dependence of the total angular momentum of the participant
system on impact parameter at different times for Pb+Pb
collisions at $\sqrt{s} = 2.76$~TeV.}
\label{fig:PbPbangmom}
\end{center}
\end{figure}

The dependence of the total angular momentum of the participant
system on impact parameter at different times is
depicted in Fig.~\ref{fig:PbPbangmom}.
The values of the angular momentum are in units of $\hbar$.
It can be seen that, similarly to the case of the total number
of participants, the total angular
momentum of the participant system increases
with time and reaches its maximum value
for each particular collision centrality at the
end of the spectator-participant partition process.

\begin{figure}
\begin{center}
\includegraphics[width=0.65\textwidth]{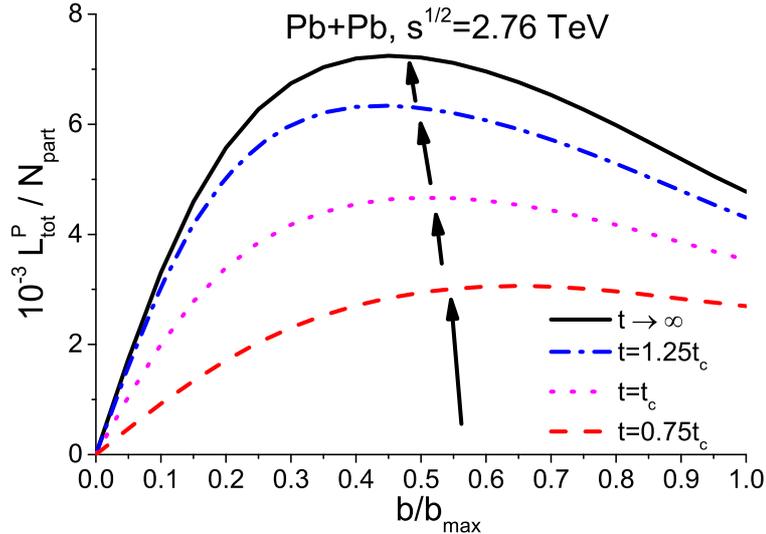}
\caption{The dependence of the participant angular momentum per
participant on impact parameter at different times for
Pb+Pb collisions at $\sqrt{s} = 2.76$~TeV.}
\label{fig:PbPbangmomoverN}
\end{center}
\end{figure}

It is also interesting to consider the time
evolution of the angular momentum of
participants per participant (per baryon charge of participants).
We note that the number of participants also changes with time.
Such a quantity contains information about an average
contribution of participant nucleons to
the total angular momentum.
The dependence of this quantity on impact parameter at different times
is depicted in Fig.~\ref{fig:PbPbangmomoverN}.
It can be seen that, similarly to the total angular momentum
of participants,
the angular momentum per participant increases with time for
any value of the impact parameter.
This means that, for any fixed value of impact parameter $b$,
the rate of increase
of the total number of participants, $N_p$, is smaller than the
rate of increase
of the total angular momentum of participants.
Another similarity is that
there is also maximum in the dependence of this quantity
on impact parameter which is shifted
in the direction of a larger $b$.
One difference is that the angular momentum per participant is
non-vanishing for large $b$,
indicating that the initial rotation
and local vorticity are significant in the range of
semi-central to even the most peripheral collisions
and needs to be accounted for.

\begin{figure}
\begin{center}
\includegraphics[width=0.65\textwidth]{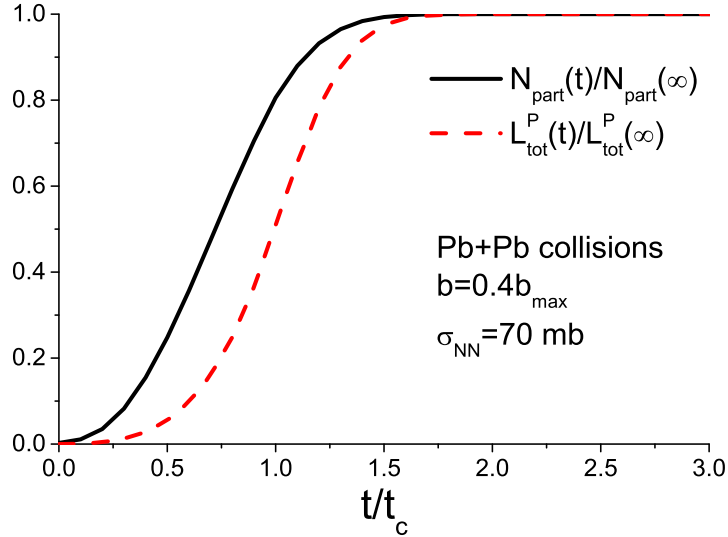}
\caption{The time dependence of the total number of
participant nucleons and of the
total angular momentum of participants divided by their final values
in Pb+Pb collisions.}
\label{fig:PbPbNvsL}
\end{center}
\end{figure}

It can be interesting to compare the rate of the increase with
time of the angular momentum of participants
with a similar rate concerning the total number
of participant nucleons. In order to do that,
we compare the time dependencies of the normalized quantities
$N_{\rm part}(t)/N_{\rm part}(\infty)$
and
$L_{\rm tot}^P(t)/L_{\rm tot}^P(\infty)$,
where
$N_{\rm part}(\infty)$ and
$L_{\rm tot}^P(\infty)$
are the values of the total number of participants
and of the total angular momentum
of participants at the end of the
spectator-participant separation stage. The time dependence of the
above-mentioned quantities is depicted in Fig.~\ref{fig:PbPbNvsL}.

It can be seen from Fig.~\ref{fig:PbPbNvsL} that
the process of increase of
the angular momentum of participants happens at a
somewhat later time in comparison to
the total number of participants,
and the most significant increase happens in time
interval $t \simeq 0.75t_c$ to $1.25t_c$.
The reason for this is that different nucleons carry different
contributions to the total participant angular momentum,
and most of the nucleons
with the largest contribution become participants
at later times,
which is also evident from the
time dependence of the angular momentum of participants related
to the number of participants (see Fig.~\ref{fig:PbPbangmomoverN}).

\subsection{Vorticity}
The classical (non-relativistic) vorticity of the participants
in the reaction plane, $(x,z)$, is defined
as
\begin{equation}
\omega_y = \omega_{xz} = - \omega_{zx}
= \frac{1}{2} \left(\partial_z v_x^P - \partial_x v_z^P \right),
\end{equation}
where $\bs v^P$ is the average 3-velocity of participants.
The emergence of the vorticity in the reaction plane
in heavy-ion collisions is attributed
to initial angular momentum of the participant system
and studies within fluid dynamical models
had shown that vorticity still remains significant
during the freeze-out stage~\cite{Csernai2013}.
Along with angular momentum such a quantity can be used
to study rotation in
the reaction plane. Another closely
related quantity is $\Lambda$ polarization which can be
detectable experimentally~\cite{Becattini2013}.
The possibility to detect rotation via differential Hanbury Brown and Twiss (HBT)
has also recently been explored~\cite{CsernaiHBT1,CsernaiHBT2}.

While in our simplified model we do not consider the subsequent evolution
of the formed participant system, most importantly the
equilibration process, we can still study the emergence of the vorticity during
the formation of this system.
To do this we assume that the transverse motion of participants
is small during the formation stage
(``no-stopping'' mode) and their average velocity can be expressed as
\begin{eqnarray}
v_x^P(t,\bs r) \approx v_y^P(t,\bs r) & \approx & 0, \\
v_z^P(t,\bs r) & \approx & v_0 \,
\frac{\rho_A^P (t, \bs r) - \rho_B^P (t, \bs r)}
{\rho_A^P (t, \bs r) + \rho_B^P (t, \bs r)}, \\
\rho_{A(B)}^P (t, \bs r)
& \approx & \rho_{A(B)}^{(0)} (t, \bs r) - \rho_{A(B)}^{S} (t, \bs r).
\label{eq:vzpart}
\end{eqnarray}
Here $\rho_{A(B)}^P (t,x,y,z)$ is the time-dependent spatial density of
participant nucleons from the target (projectile).
For the relativistic case we follow the definition
from Ref.~\cite{Molnar2010}
\begin{equation}
\omega_{\nu}^{\mu}
= \frac{1}{2} \left( \nabla_{\nu} u^{\mu} - \nabla^{\mu} u_{\nu} \right),
\end{equation}
where $u^{\mu} = \gamma (1, \bs v)$,
$\nabla_{\alpha} = \Delta_{\alpha}^{\beta} \partial_{\beta}$
and $\Delta^{\mu\nu} = g^{\mu\nu} - u^{\mu} u^{\nu}$.
Similarly to Ref.~\cite{Csernai2013} we
neglect the collective acceleration in comparison with rotation,
i.e., $|\partial_{\tau} u^{\mu}| \ll | \partial_x u^z |$, and
get the following expression for the relativistic
vorticity $\omega_z^x$ in the reaction plane
\begin{equation}
\omega_z^x = -\omega_x^z
= -\frac{1}{2} \gamma \partial_x v_z - \frac{1}{2} v_z \partial_x \gamma,
\end{equation}
where $\gamma = (1-v_z^2)^{-1/2}$.
Here we already take into account that $v_x = v_y = 0$ in our model.

\begin{figure}
\begin{minipage}{.48\textwidth}
\centering
\includegraphics[width=\textwidth]{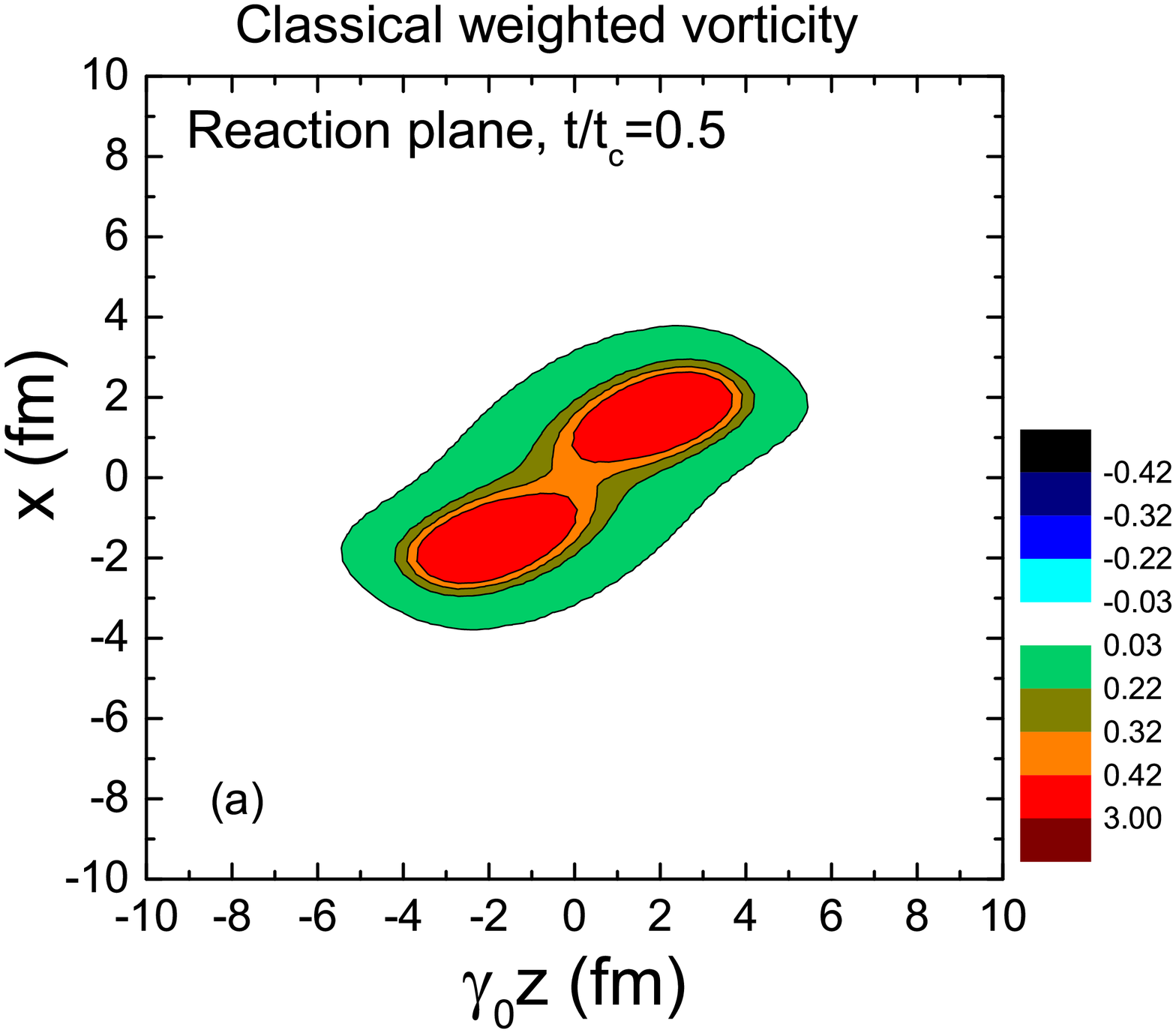}
\end{minipage}
\begin{minipage}{.48\textwidth}
\includegraphics[width=\textwidth]{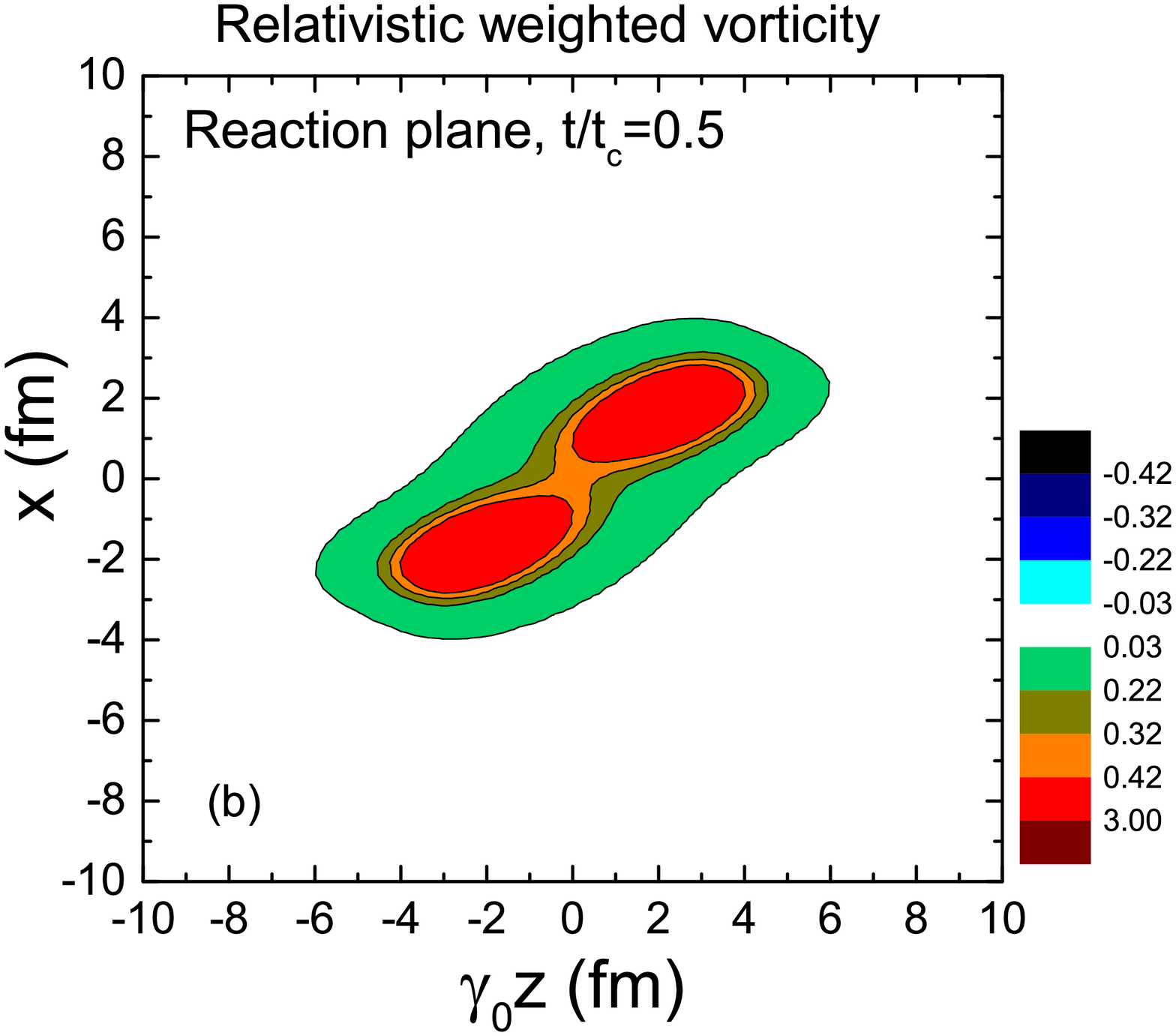}
\end{minipage}
\caption{The (a) classical and (b) relativistic weighted
participant vorticity, $\Omega_{zx}$, in units of $c$/fm, calculated
in the reaction plane, i.e. $(xz)$ plane, at time moment $t=0.5t_c$
in Pb+Pb collisions.
The collision energy is $\sqrt{s_{_{NN}}} = 2.76$~TeV and $b=0.7b_{\rm max}$.
The collision axis $z$ is scaled with
$\gamma$-factor $\gamma_0$, which corresponds
to the collision energy.}
\label{fig:vort1}
\end{figure}

\begin{figure}
\begin{minipage}{.48\textwidth}
\centering
\includegraphics[width=\textwidth]{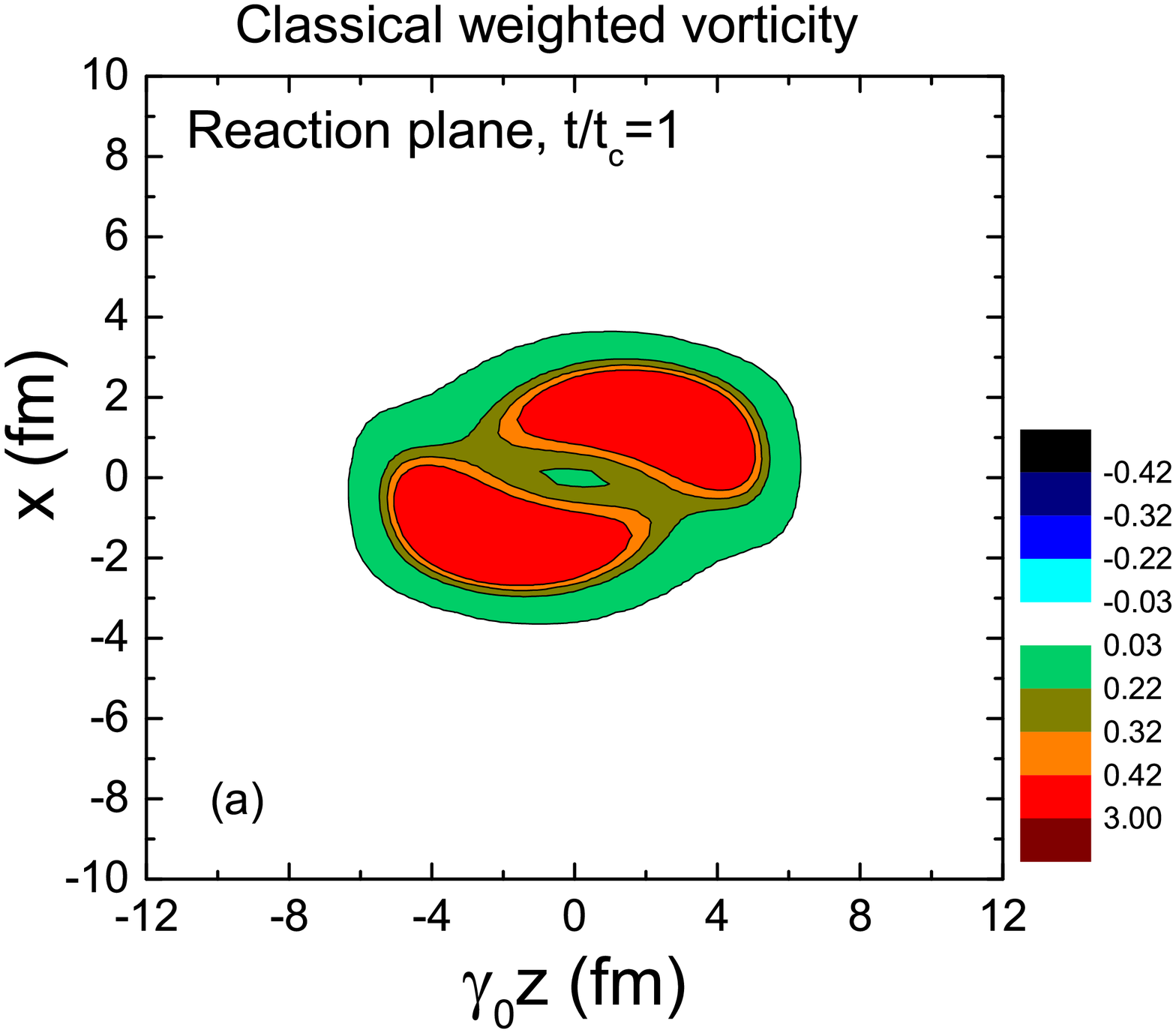}
\end{minipage}
\begin{minipage}{.48\textwidth}
\includegraphics[width=\textwidth]{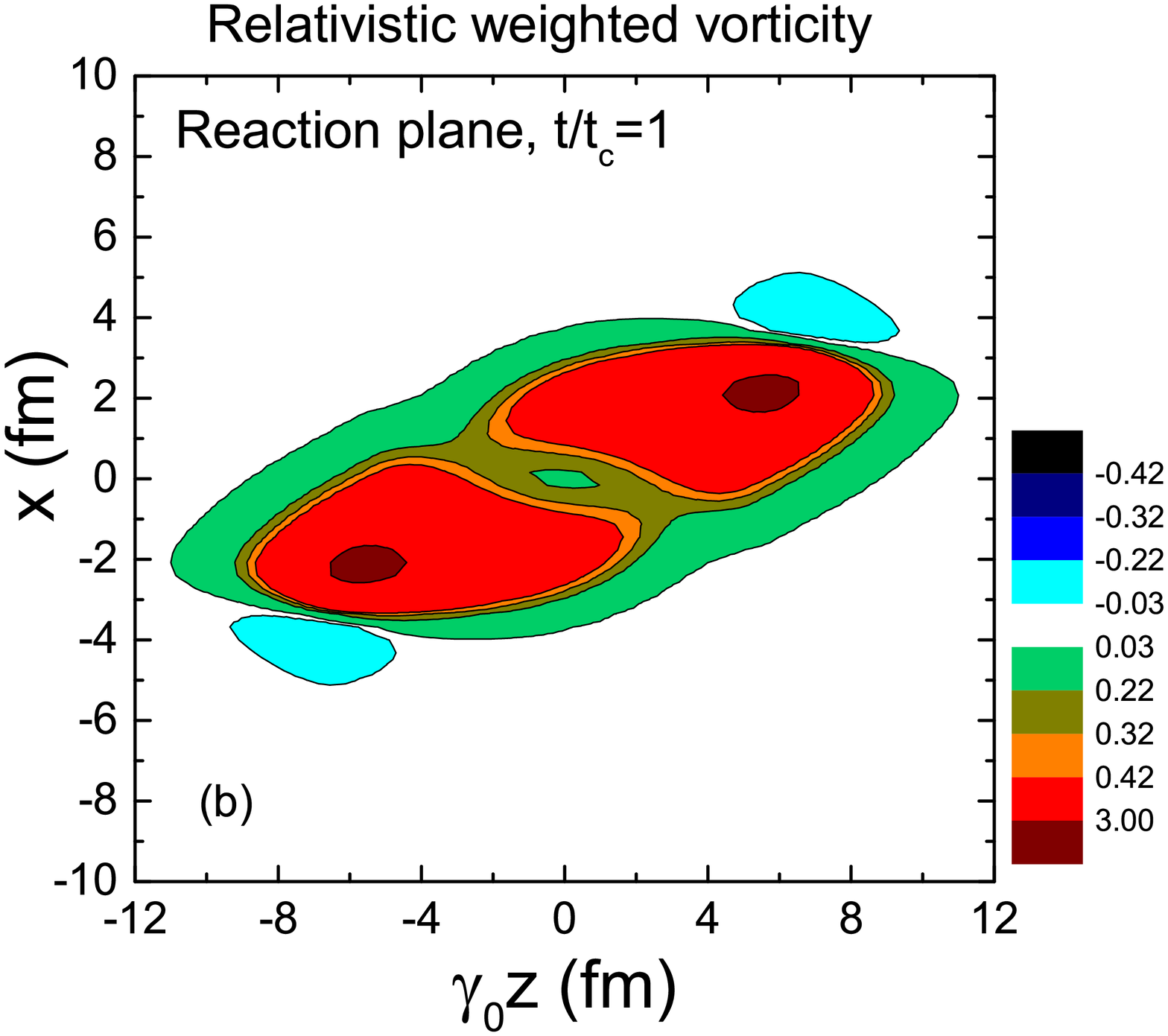}
\end{minipage}
\caption{Same as Fig.~\ref{fig:vort1}, but for $t=t_c$.}
\label{fig:vort2}
\end{figure}

\begin{figure}
\begin{minipage}{.48\textwidth}
\centering
\includegraphics[width=\textwidth]{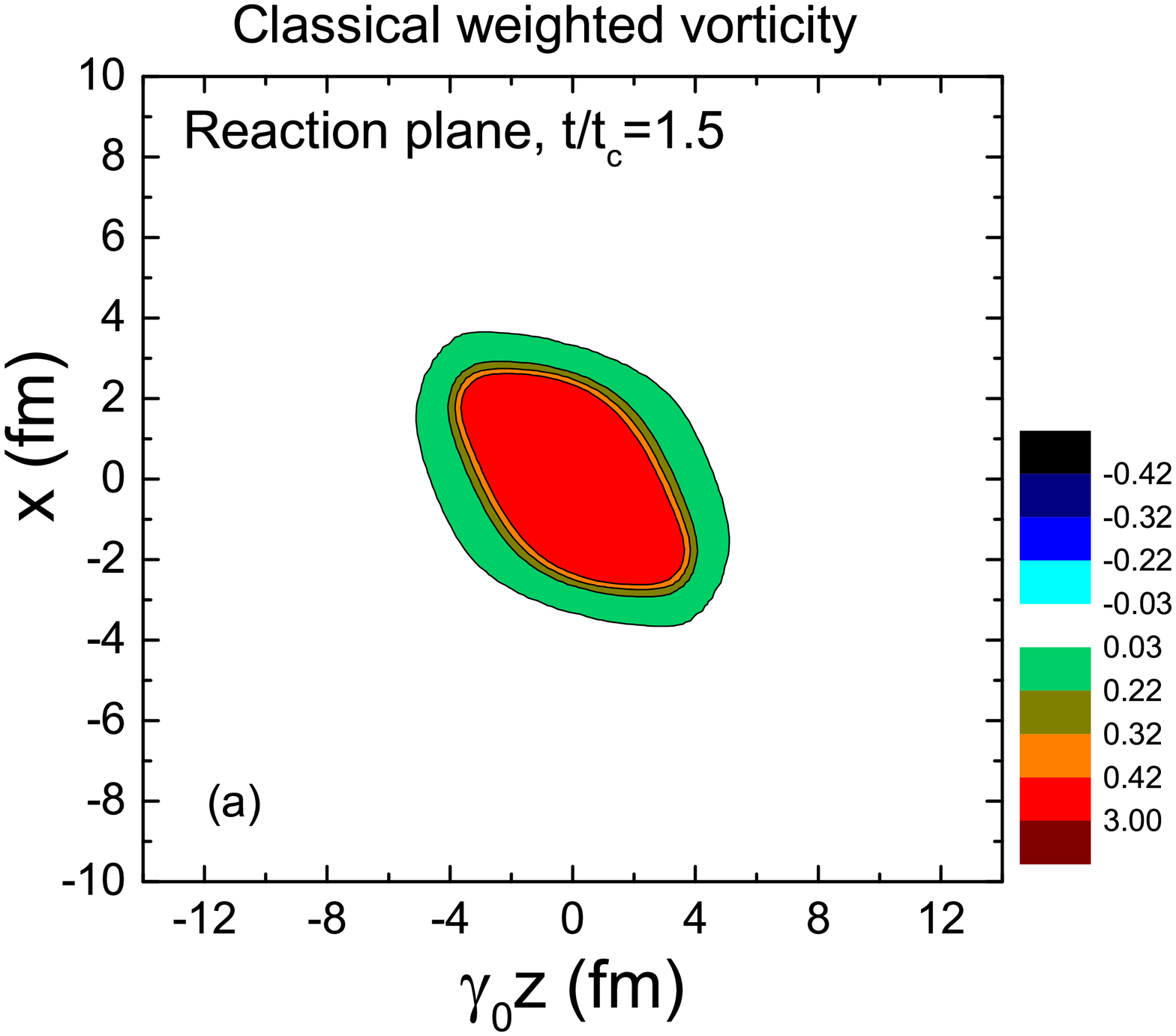}
\end{minipage}
\begin{minipage}{.48\textwidth}
\includegraphics[width=\textwidth]{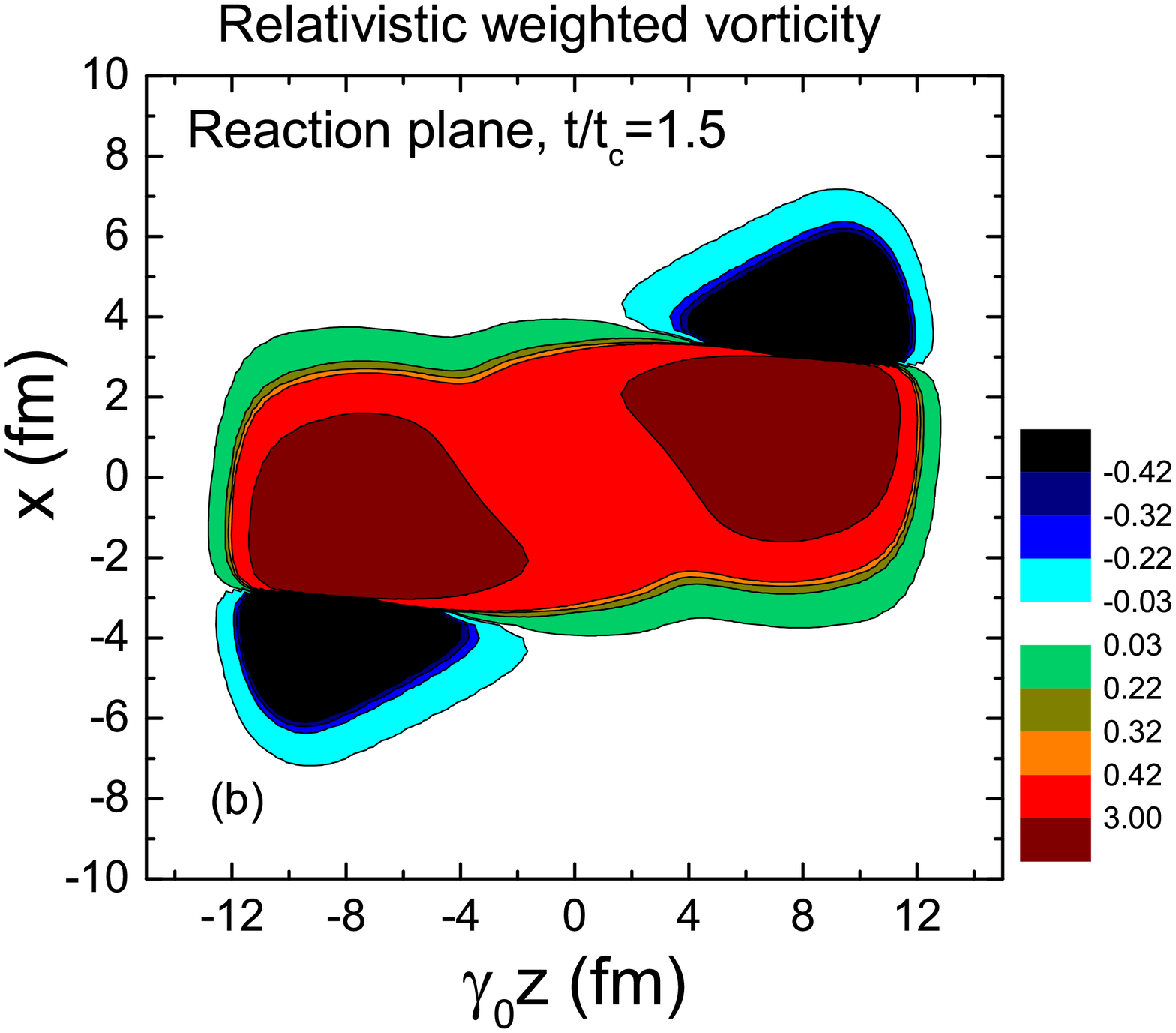}
\end{minipage}
\caption{Same as Fig.~\ref{fig:vort1}, but for $t=1.5t_c$.}
\label{fig:vort3}
\end{figure}

Similarly to Ref.~\cite{Csernai2013},
we also use the
weights proportional to the energy density to better reflect
the collective dynamics. The energy-density weighted vorticity
for both classical and relativistic cases is then
\begin{equation}
\Omega_{zx} = w(t, x, z) \omega_{zx},
\end{equation}
where the weight, $w(t, x, z)$, is
\begin{equation}
w(t, x, z)
= \frac{\epsilon^P(t,x,y=0,z)}{\langle \epsilon^P(t,x,y=0,z) \rangle}.
\end{equation}
Here, $\epsilon^P(t,x,y,z)
= \displaystyle \frac{\sqrt{s}}{2} \left( \rho_A^P + \rho_B^P \right)$
is the
energy density of the participants and
$\langle \epsilon^P(t,x,y=0,z) \rangle$
is the average energy density in the reaction plane at time $t$.
For averaging we use the region
$-1.5R_0 < x < 1.5R_0, \, -1.5R_0 < \gamma_0 z < 1.5R_0$.
Results of the calculations of the classical and relativistic weighted
vorticity in the reaction plane at different time moments are presented
in Figs.~\ref{fig:vort1}-\ref{fig:vort3}.

The presented results illustrate the emergence of rotation
during the formation of the participant system. Also, it is seen
that there may exist substantial differences in results when using different
definitions of vorticity indicating that its relativistic generalization
is not trivial. It should also be noted, however, that the proposed model
does not describe the evolution of the participant system after its formation
and using the ``no-stopping'' assumption,
Eq. \eqref{eq:vzpart}, allows us to only
give qualitative rather than quantitative picture,
especially for times $t>t_c$.

\section{Conclusions}
\label{sec:conclusions}

The identification of different stages of the initial state
is important if we want to discuss the results of multimodule
models or hybrid models. While the middle part of a heavy ion reaction
is usually well described by the fluid dynamical model, different
initial states and different final-state approximations are used
in such kinds of combined models.

In the Particle in Cell relativistic (PICR) fluid dynamical model \cite{hydro1,hydro2} the initial
state assumes a dynamical evolution in a Yang-Mills field theoretical
model \cite{M1,M2}, which has some features similar to the model
presented here. The time when the PICR calculation starts corresponds
to a configuration when the two nuclei have interpenetrated each
other and were near to be stopped by the Yang-Mills field. In
the timescale of this model this configuration corresponds
to a time moment not earlier than $2 t_c$.
The subsequent (3+1)-dimensional fluid dynamical development
led to increased rotation due to the Kelvin-Helmholtz Instability
(KHI) in certain favorable configurations.
The initial time moment
of the hydrodynamical evolution in the hybrid approach
based on UrQMD model~\cite{Petersen2008} is also closely
related to the temporal scale $t_c$ of our model.
There $2t_c$ is assumed to be the earliest possible thermalization time
and, consequently, the earliest possible initial time moment
of the hydrodynamical
evolution, which should not be smaller than 1~fm/$c$.

The present model is based on a conserved nucleon picture.
For example, the angular momentum per nucleon assumes conserved nucleons.
At very high energies numerous hadron pairs are created including
baryon pairs, so the concept of the model should be implemented
for the conserved baryon charge.

Physically, the prehydrodynamical stage will remain nearly the
same; however, the high parton density may influence the dynamics
already after $t_c$. Especially, collective force fields may change
the dynamics, and may speed up equilibration, which then leads to
collective effects like the KHI.

The vorticity characteristics shown in Fig. \ref{fig:vort3}
are interesting. The participant domain has substantial positive
vorticity. This agrees well with the fluid dynamical calculations.
The spectators show negative vorticity, this is arising from the
particle loss due to collisions from the spectator domain.
Because the spectators are not considered at all in the PICR
calculations this effect is not covered by these model calculations.

Notice the large difference between the non-relativistic and
relativistic vorticities in Figs. \ref{fig:vort2} and \ref{fig:vort3}.
This is due to the relativistic $\gamma$ factors, which are large
in the present calculation as there are only collisions, no
collective forces or pressure. In the PICR calculations these
collective interactions decrease velocity differences both in the
initial state model and in the fluid dynamics, thus the difference
between the  non-relativistic and relativistic vorticities is modest.

The initial state model in the PICR calculations is dominated by attractive
collective Yang-Mills fields, which keep the system more compact and uniform.
Some versions of the Color-glass Condensate (CGC) initial state models
have similar features.
Also in the PICR model sharp initial nuclear surfaces are assumed
instead of Woods-Saxon surface profiles. This makes the typical times
$t_c$ and $2t_c$ shorter.
On the other hand for molecular dynamics models
(or to some extent for hybrid models)
with MC-Glauber initialization the present model provides a good estimate
for the initial times.  See Appendix A.

The formation of a quark-gluon plasma (QGP) leads to more rapid equilibration and to critical
fluctuations. These also facilitate the equilibration of rotation especially
in low viscosity fluid dynamical models like PICR with KHI. Before the
final hadronization the perturbative vacuum may keep the participant system
more compact and then rapid hadronization from a supercooled QGP has the
best chances to show observable signs of rotation at the final freeze out.
To detect the observable signs of Global Collective Flow patterns these should
be separated from random fluctuations as described in Ref. \cite{CS14}.

At the same time, for the development of the initial rotation and
vorticity the present model provides an excellent guidance for all
dynamical models of peripheral heavy ion reactions.

\begin{acknowledgments}
The work of D.A. was supported  by the Program of Fundamental Research of the
Department of Physics and Astronomy of NAS and by the State Agency of Science,
Innovations and Informatization of Ukraine contract F58/384-2013.
\end{acknowledgments}

\newpage

\section*{Appendix}

\subsection{Reaction density of binary collisions}
\label{app:Gtz}
Our model gives the possibility to calculate the density, $\Gamma(t,\bs r)$,
of binary collisions between nucleons from colliding nuclei, which
describes the number of binary reactions per unit volume per
unit time.
Since these binary collisions are beam directed, the  
relative velocity of nucleons is $2v_0$.
Exploiting this and taking into account the ballistic distribution functions
$f^{(0)}_{A(B)}$ of the colliding nucleons
[see Eq.~\eqref{eq:fAB_final}]
one can write down the four-density of binary reactions as
\begin{equation}
\Gamma_{\rm coll} (t, \bs r) = \sigma_{_{NN}} \, 2 v_0 \,
\rho_A^{(0)} (t, \bs r) \, \rho_B^{(0)} (t, \bs r).
\end{equation}
The total average number of binary collisions $N_{\rm coll}$ is
\begin{eqnarray}
N_{\rm coll} & = & \int dt\, d \bs r\, \Gamma_{\rm coll} (t, \bs r)
\\
&& \hspace{-10mm} =  \sigma_{_{NN}} \, 2v_0 \gamma_0^2 \int dt\, d \bs r\,
\rho_{_{WS}} \big(x-b/2,y,\gamma_0[z - v_0 (t - t_c)]\big) \,
\rho_{_{WS}} \big(x+b/2,y,\gamma_0[z + v_0 (t - t_c)]\big)\,.
\nonumber
\end{eqnarray}
Making a change of variables $(t,z) \to (z_1, z_2)$ as $z_1 =
\gamma_0[z - v_0 (t - t_c)]$, $z_2 = \gamma_0[z + v_0 (t - t_c)]$ we get
\begin{eqnarray}
N_{\rm coll} & = & \sigma_{_{NN}} \int d \bs r_{\perp} \int d z_1\,
\rho_{_{WS}} \big(x-b/2, y, z_1\big) \int d z_2\, \rho_{_{WS}} \big(x+b/2, y, z_2\big)
\nonumber \\
& = & \sigma_{_{NN}}
\int dx dy \, T_A\big(x-b/2,y\big) \, T_B\big(x+b/2,y\big)\, =\, \sigma_{_{NN}} A^2\, t(b),
\label{eq:Nbinary}
\end{eqnarray}
where $t(b)$ is the nuclear overlap function, normalized to unity, which depends
on the impact parameter.
Equation \eqref{eq:Nbinary} coincides with the expression for average number of
binary collisions in the analytical Glauber model.
Our model, however, allows one to study also the temporal and longitudinal
structure of the binary collisions.

\begin{figure}
\begin{center}
\includegraphics[width=0.65\textwidth]{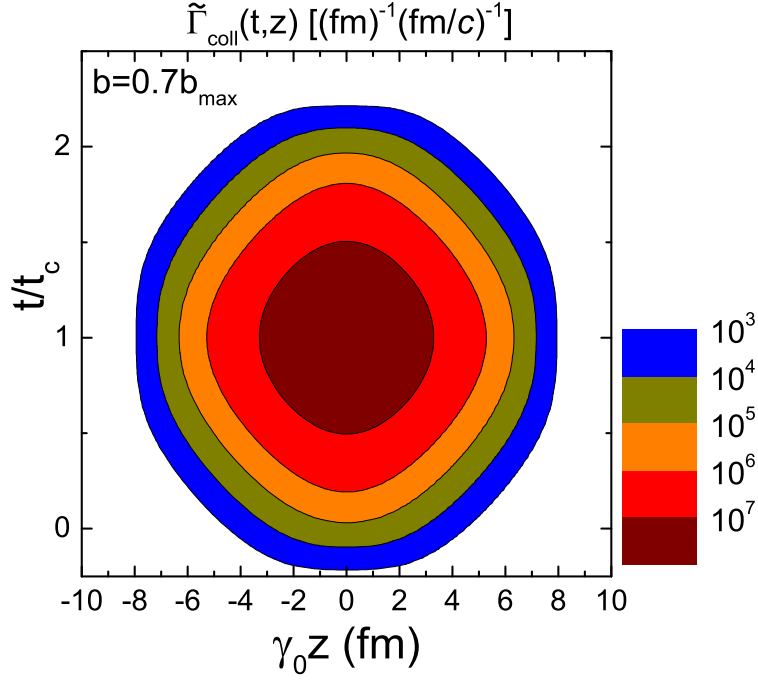}
\caption{The binary reaction density
$\tilde{\Gamma}_{\rm coll} (t, z)$
in coordinates $(t,z)$ in Pb+Pb collisions.
The collision energy is
$\sqrt{s_{_{NN}}} = 2.76$~TeV and $b=0.7b_{\rm max}$.
The collision axis, $z$, is scaled with $\gamma$-factor,
$\gamma_0$, which corresponds to the collision energy.}
\label{fig:G0tz}
\end{center}
\end{figure}

Let us consider the quantity
$\tilde{\Gamma}_{\rm coll} (t, z) = \int dx dy \Gamma_{\rm coll}(t, \bs r)$,
which represents the two-dimensional space-time structure of the binary collisions.
This quantity is depicted in Fig.~\ref{fig:G0tz}.
It is instructive to compare the structure of two-dimensional binary collisions given in
Fig.~\ref{fig:G0tz} with space-time reaction zones which were investigated in
Ref.~\cite{Anchishkin2013} exploiting UrQMD: very similar features of the distribution of
collisions can be immediately found at earlier times.
Besides, it is explicitly seen in Fig.~\ref{fig:G0tz} how natural and useful for
the description of the initial stage is the time
scale $t_c$, which is a unit of a measuring the time axis.

\begin{figure}
\begin{center}
\includegraphics[width=0.65\textwidth]{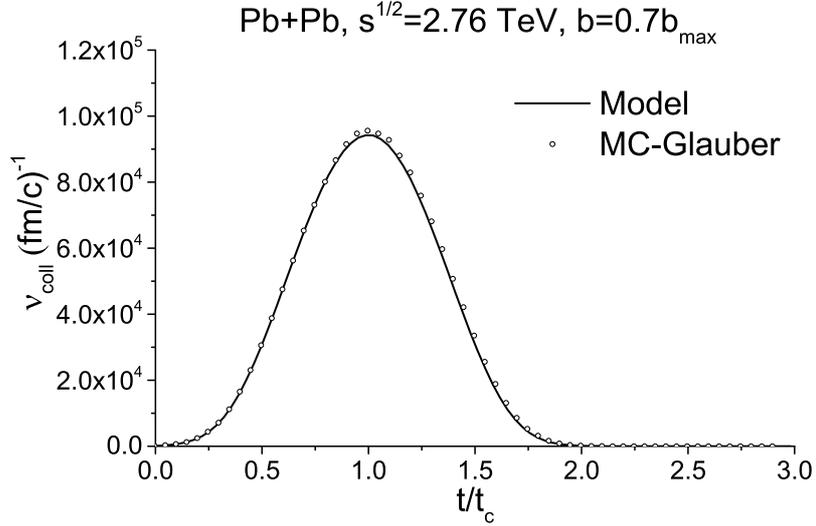}
\caption{The frequency of binary collisions
$\nu_{\rm coll} (t)$ in Pb+Pb collisions
calculated in our model and in Glauber Monte Carlo.
The collision energy is
$\sqrt{s_{_{NN}}} = 2.76$~TeV and $b=0.7b_{\rm max}$.}
\label{fig:nu0}
\end{center}
\end{figure}
It is useful to make a comparison of our model
to MC-Glauber.
In MC-Glauber one can take into account
correlations generated by the collision mechanism (dubbed ``twin'' correlations
in Ref.~\cite{Blaizot2014}), i.e.
that nucleons can only collide if they are close by in the transverse plane.
In order to make a comparison
we consider the
frequency of binary reactions,
$\nu_{\rm coll} (t) = \int dz \tilde{\Gamma}_{\rm coll} (t, z)$,
which can be calculated in our model and also in MC-Glauber.
To calculate this quantity in MC-Glauber we follow
the usual procedure, recently
described in Ref. \cite{ALICECentrality2013}, but
also add additional step to determine time dependence:
\begin{enumerate}
\item We generate the initial positions of nucleons in colliding nuclei by
using the Woods-Saxon distribution with the same parameters that are used
in our analytical model.
\item We consider all possible binary collisions between the nucleons from different
colliding nuclei by calculating the distance, $d_{\rm trans}$, between
them in the transverse plane.
In case it satisfies the inequality
$\displaystyle d_{\rm trans} < \sqrt{\frac{\sigma_{_{NN}}}{\pi}}$,
we register a binary collision.
\item We calculate the time moment for each binary collision as $\displaystyle
t = \frac{|z_1-z_2|}{2 v_0}$, where $z_1$ and $z_2$ are the longitudinal coordinates
of the two colliding nucleons in the collider center-of-mass frame at $t=0$.
\end{enumerate}
The frequency of binary reactions calculated in our analytical model and in the
MC-Glauber are depicted in Fig.~\ref{fig:nu0}.
It is seen that both graphs virtually coincide, which further indicates that our
model is consistent with Glauber approach and also that event-by-event fluctuations
and ``twin'' correlations
have negligible effect on a frequency of the binary reactions.


\end{document}